\documentclass[10pt,journal]{IEEEtran}
\IEEEoverridecommandlockouts
\usepackage{algorithmicx,algorithm}
\usepackage{algpseudocode}
\usepackage{cite}
\usepackage[utf8x]{inputenc} 
\usepackage{amsmath}
\usepackage{amssymb,amsmath}
\usepackage{graphicx}
\usepackage{epsfig}
\usepackage{grffile}
\usepackage{balance,color}
\usepackage{caption}
\usepackage{subcaption}
\usepackage{xcolor, soul}
\sethlcolor{cyan}
\usepackage{booktabs}
\usepackage{multirow}
\usepackage{makecell}
\providecommand{\keywords}[1]{{\textit{Index Terms}}}

\begin{document}

\title{ RIoT Digital Twin: Modeling, Deployment, and Optimization of Reconfigurable IoT System with Optical–Radio Wireless Integration
} 
\author{Alaa Awad Abdellatif$^\star$, Sérgio Silva$^\star$, Eduardo Baltazar$^\star$, Bruno Oliveira$^\star$, Senhui Qiu$^+$, Mohammud J. Bocus$^+$, Kerstin Eder$^+$, Robert J. Piechocki$^+$, Nuno T. Almeida$^\star$, Helder Fontes$^\star$  \\
\begin{tabular}{c}
 $^\star$ INESC TEC and Faculdade de Engenharia, Universidade do Porto, Portugal  \\  
 $^+$ School of Computer Science, University of Bristol, Bristol, UK  \\  
		\thanks { This study is supported by the SUPERIOT project, funded by the Smart Networks and Services Joint Undertaking (SNS JU) under the European Union's Horizon Europe research and innovation programme (Grant Agreement No 101096021), with additional funding from UK Research and Innovation (UKRI) under the UK government’s Horizon Europe funding guarantee.  }  
\end{tabular}
}
\maketitle

\begin{abstract}
This paper proposes an optimized Reconfigurable Internet of Things (RIoT) framework that integrates optical and radio wireless technologies with a focus on energy efficiency, scalability, and adaptability. To address the inherent complexity of hybrid optical–radio environments, a high-fidelity Digital Twin (DT) is developed within the Network Simulator 3 (NS-3) platform. The DT models deploy subsystems of the RIoT architecture, including radio frequency (RF) communication, optical wireless communication (OWC), and energy harvesting and consumption mechanisms that enable autonomous operation. Real-time energy and power measurements from target hardware platforms are also incorporated to ensure accurate representation of physical behavior and enable runtime analysis and optimization. 
Building on this foundation, a proactive cross-layer optimization strategy is devised to balance energy efficiency and quality of service (QoS). The strategy dynamically reconfigures RIoT nodes by adapting transmission rates, wake/sleep scheduling, and access technology selection. Results demonstrate that the proposed framework, combining digital twin technology, hybrid optical–radio integration, and data-driven energy modeling, substantially enhances the performance, resilience, and sustainability of 6G IoT networks. 


\end{abstract}

\section{Introduction   \label{sec:Introduction} }

The rapid growth of the Internet of Things (IoT) has accelerated the need for network solutions that are energy-efficient, scalable, and adaptable to diverse application requirements. With billions of connected devices transforming healthcare, transportation, smart cities, and environmental monitoring \cite{cisco2018internetreport,9866918}, IoT is becoming a core component of modern digital infrastructure. However, this massive growth raises critical sustainability concerns. The manufacture, deployment, and operation of billions of devices significantly increase global energy consumption, resource usage, and electronic waste. 
From a sustainability perspective, two main aspects must be addressed: (a) the energy consumption of IoT nodes, and (b) the sustainability of the networks interconnecting them. Although individual IoT nodes consume minimal power, their cumulative impact is substantial at a global scale. Most IoT nodes are battery-powered or rely on energy harvesting and are expected to function for long periods without maintenance, making energy efficiency essential. Furthermore, IoT connectivity relies on network infrastructures that demand energy, physical space, and maintenance. This challenge is further amplified by the vision of 6G, which calls for ultra-dense connectivity, near-zero energy waste, and environmentally sustainable IoT infrastructures \cite{rashid2025review}. Therefore, sustainable IoT solutions must optimize device energy use while leveraging existing network infrastructure to reduce environmental and operational costs. 

While radio technologies remain the backbone of IoT connectivity, they face growing challenges related to spectrum scarcity, interference, and energy consumption. Optical Wireless Communication (OWC) has emerged as a promising complement to radio-based systems, offering wide unlicensed bandwidth, inherent physical-layer security, and seamless integration with existing lighting infrastructure \cite{9771322}. However, its performance remains highly dependent on line-of-sight (LoS) conditions and is sensitive to environmental factors such as weather, mobility, and obstruction. 
Integrating radio and optical communications within a unified, reconfigurable framework enables the exploitation of their complementary strengths, combining the ubiquity and robustness of radio with the efficiency and security of optical links. However, hybrid RF–OWC networks add complexity, as nodes must intelligently select between modalities before associating with an access point. This results in a multidimensional, context-aware decision problem that jointly considers network characteristics, energy efficiency, mobility, and channel dynamics. Furthermore, in such a dynamic environment, real-world experimentation becomes costly and time-consuming, highlighting the need for a Digital Twin (DT) to enable accurate modeling, safe testing, and data-driven optimization of hybrid IoT networks. 

\begin{figure}[t!]
	\centering
		\scalebox{1.8}{\includegraphics[width=0.34 \textwidth]{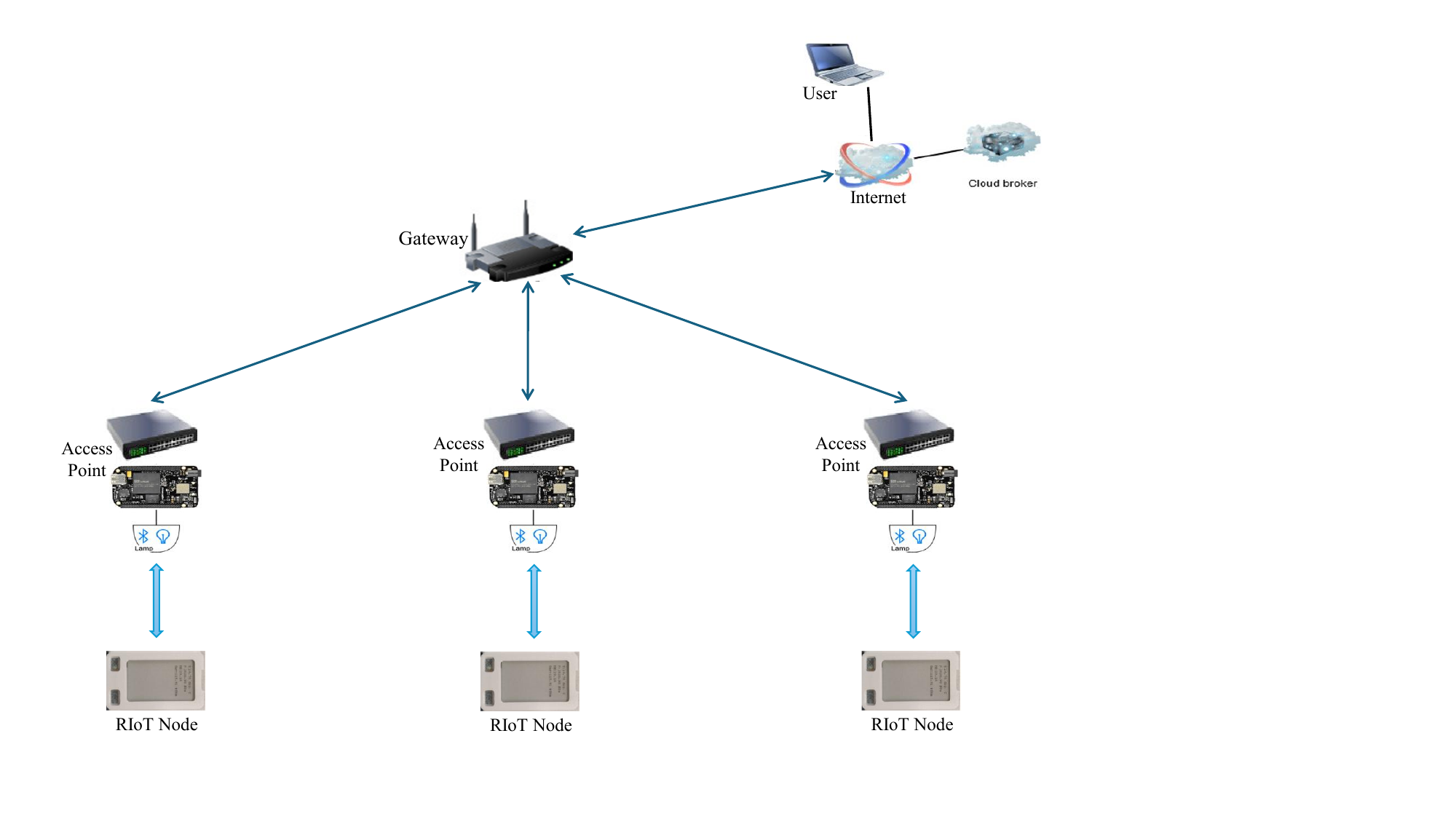}}
	\caption{ The considered RIoT system.   } 
	\label{fig:system}
\end{figure}

In light of these challenges, this paper introduces a Reconfigurable IoT (RIoT) framework that integrates hybrid RF/OWC communication, energy-aware modeling, and DT technology to enable scalable, sustainable, and context-aware 6G IoT networks. As illustrated in Fig. \ref{fig:system}, the proposed system leverages reconfigurable IoT nodes, multimodal connectivity, and intelligent energy management to optimize power consumption while satisfying application-specific QoS requirements. This RIoT framework advances the state of the art by embedding sustainability into system design and operation, offering flexibility, adaptability, and long-term resilience in highly dynamic environments \cite{10118842, 9144301}.   
Within this framework, a DT is developed to accurately represent RIoT nodes and their key subsystems, including RF and OWC communication modules as well as energy harvesting and consumption behavior. By incorporating hardware-based energy measurements, the DT ensures high fidelity between physical and virtual environments, enabling reliable real-time monitoring, analysis, and optimization. Building on this foundation, a proactive cross-layer optimization strategy is introduced to dynamically adapt communication modalities, access technologies, and operational parameters based on traffic demands and network conditions. This approach minimizes energy consumption while maintaining application-level QoS, thereby addressing the multi-dimensional decision-making challenges inherent in hybrid RF/OWC IoT networks. 

To this end, the proposed framework pursues three main contributions: 
\begin{enumerate}
    \item \textbf{Energy Modeling and Prediction:} Develop energy consumption models grounded in hardware-level measurements, and validate them through empirical data. These models capture the energy dynamics of individual network elements as well as the network as a whole. Leveraging these models, we analyze energy consumption patterns, supported by high-level simulations of RIoT node abstractions. 
    \item \textbf{Digital Twin Development:} Design and implement a DT of RIoT nodes to enable scalable and energy-efficient solutions that are adaptable to varying operational demands. The DT serves as a virtual replica of the physical hybrid network, allowing real-time monitoring, analysis, and optimization of node behavior.  
    \item \textbf{Proactive Cross-Layer Optimization:} Establish a dynamic mechanism to optimize communication modality and access technology configurations in real time. Through integration with the DT, the proposed framework enables proactive reconfiguration of RIoT nodes by adaptively adjusting communication modalities, access technologies, and operational parameters. This approach maximizes energy efficiency while ensuring application-level QoS is consistently maintained. 
\end{enumerate}

The remainder of this paper is organized as follows. Section~\ref{sec:Related} reviews related work on hybrid optical–radio IoT systems and the use of ns-3 for modeling and evaluation. Section~\ref{sec:DT_NS3} describes the digital abstraction of the RIoT node and its implementation within the NS-3 environment. Section~\ref{sec:Modeling} presents the energy measurement setup and calibration of DT models, including node-level and access point energy consumption. Section~\ref{sec:Optimization} introduces the proposed low-complexity optimization algorithm for adaptive node configuration. Section~\ref{sec:Evaluation} discusses the experimental and DT-based performance results. Finally, Section~\ref{sec:conclusion} concludes the paper.  

\section{Related Work  \label{sec:Related} }

The rapid growth of IoT applications has driven extensive research into energy efficiency, scalability, communication technologies, and system sustainability. Several related works have investigated these aspects across domains such as healthcare, smart cities, environmental monitoring, transportation, and industrial automation, emphasizing both the societal impact of IoT and the challenges associated with large-scale deployment \cite{11121897, 10768987, 9866918}. 

Given this wide scope, this section focuses on two main related areas: (i) research on integrating optical and radio wireless communications in IoT systems, and (ii) prior efforts on using the NS-3 network simulator for modeling, analysis, and evaluation of IoT systems. 

\subsection{Optical–Radio Wireless Integration} 

RF technologies such as Wi-Fi, Bluetooth, and LoRa have traditionally dominated IoT connectivity. However, OWC has gained increasing attention as a complementary alternative due to its large unlicensed spectrum and inherent physical-layer security. OWC includes long-range Free Space Optics (FSO) and short-range Visible Light Communication (VLC) and LiFi, which reuse existing lighting infrastructure to provide high-capacity wireless access. Recent advancements in OWC have demonstrated data rates in the order of hundreds of Gb/s using advanced LED and laser-based transmitters \cite{11153838}. However, OWC remains constrained by line-of-sight requirements and limited coverage. To address these limitations, hybrid RF/OWC architectures have been proposed, combining the robustness and mobility support of RF with the high capacity and energy efficiency of OWC \cite{abuella2021hybrid, 9743352, 10511290}. Such hybrid systems are particularly relevant for IoT networks, where dynamic environments, energy constraints, and heterogeneous traffic demands necessitate adaptive and context-aware communication mechanisms \cite{9780604}.  However, these studies stop short of jointly addressing: (i) hybrid RF–OWC integration with node reconfiguration in large-scale IoT, (ii) real-time energy-harvesting and consumption modelling for sustainability, and (iii) DT implementations for real-time adaptive control in such environments.

While hybrid RF/OWC architectures have been explored to enhance coverage and capacity in IoT networks, existing approaches do not fully address real-time, intelligent selection of communication modalities and access points under varying network, energy, and environmental conditions. Traditional systems like WLANs perform access point (AP) selection based mainly on signal strength, interference, or load \cite{9726129}, whereas hybrid RF/OWC networks introduce a more complex decision process, requiring nodes to first choose between RF and OWC, then associate with a suitable AP. This selection must account for heterogeneous AP capabilities, coverage, reliability, environmental sensitivity, asymmetric uplink/downlink characteristics, and the presence of dual-mode devices within overlapping service areas \cite{8917797}. 
As a result, modality and AP selection become a high-dimensional, context-aware optimization problem involving data rates, energy consumption, QoS requirements, security, mobility, deployment costs, and rapidly varying channel conditions. 
For instance, the work in \cite{10899885} introduces empirical energy modeling and adaptive parameter tuning for RF-based IoT devices, demonstrating the importance of data-driven energy optimization. However, its scope is limited to RF communication and node-level adjustments. It does not address hybrid RF–OWC integration, real-time DT synchronization, or large-scale network reconfiguration across heterogeneous access technologies. 
Therefore, scalable and high-fidelity models are required to jointly capture physical-layer behavior, device-level energy profiles, and environmental dynamics, enabling accurate energy-aware optimization. This motivates the adoption of DT technology, which enables adaptive and predictive strategies for modality selection and RIoT node reconfiguration while balancing energy efficiency and QoS, capabilities that are explored in detail in the following sections. 


\subsection{Simulation of RIoT Nodes in NS-3}

Developing a DT of a RIoT node requires accurate modeling of its key subsystems, including RF and OWC communication modules, as well as energy harvesting and consumption components. Bluetooth Low Energy (BLE) has been widely used in low-power IoT systems for RF communication due to its configurable trade-offs between latency, throughput, and energy consumption~\cite{11}. However, NS-3 does not provide native BLE support. The most commonly used solution is a community-developed module~\cite{12}, which models the BLE PHY/MAC layers based on Bluetooth 4.2 and allows configuration of key parameters such as connection interval and slave latency—essential for energy optimization. Despite its usefulness, this module lacks higher-layer protocol support, does not implement GFSK modulation, and is poorly documented, making integration and debugging difficult.

OWC, especially Visible Light Communication (VLC), has gained attention as a complement to RF technologies due to its high bandwidth, directional transmission, and immunity to electromagnetic interference. Its suitability for IoT applications is further supported by the widespread availability of LED lighting infrastructure that can be reused for data transmission~\cite{17,18}. In NS-3, however, VLC support remains limited \cite{19, 21}. One notable contribution is the hybrid Wi-Fi/VLC module proposed in~\cite{19}, which includes SNR and BER models and supports basic modulation schemes such as OOK and VPPM. Nonetheless, this implementation is constrained by its inheritance from Point-to-Point NetDevice, lack of PHY/MAC separation, and absence of a PHY state machine, making it unsuitable for wireless, multi-hop, or energy-aware simulations. 
Furthermore, while NS-3 includes an energy framework with sources, harvesters, and device energy models (DEMs), existing models are tailored to conventional RF devices and do not capture the energy behavior of hybrid BLE/VLC communication or peripheral tasks such as localization and display updates \cite{21}. As a result, current simulations overlook realistic energy dynamics and limit cross-layer optimization. To address this gap, we develop custom DEMs based on real hardware measurements, enabling accurate energy-aware DT modeling and adaptive reconfiguration of RIoT nodes. 



\section{Design and Implementation of the NS-3 RIoT Node \label{sec:DT_NS3}}

The DT of the RIoT node in NS-3 is developed as a high-fidelity simulation framework for evaluating and optimizing sustainable IoT networks. This section describes the digital abstraction of the RIoT node and its implementation within the NS-3 environment.

\subsection{ RIoT Digital Twin Architecture}  

\begin{figure}[t!]
	\centering
		\scalebox{1.34}{\includegraphics[width=0.34 \textwidth]{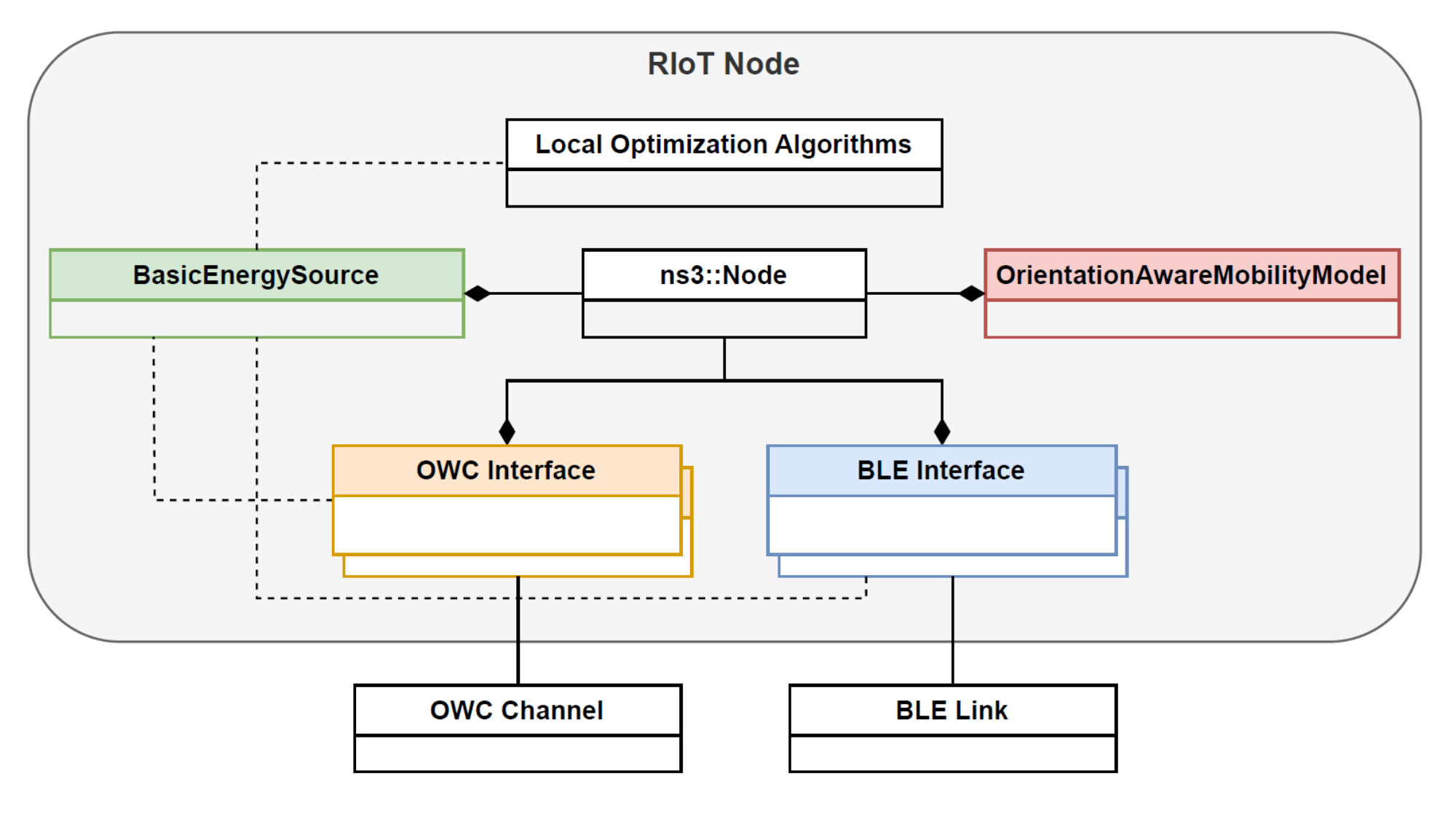} } 
	\caption{ The developed architecture of the RIoT node in NS-3.  } 
	\label{fig:architecture}
\end{figure}

The {RIoT node architecture}, illustrated in Figure~\ref{fig:architecture}, is developed by translating the physical node's structural and functional characteristics into corresponding {NS-3} components. 
This digital representation accurately models the node's primary subsystems, including: 
(i) the {RF interface}, implemented through Bluetooth Low Energy (BLE) communication; 
(ii) the {optical interface}, represented by Visible Light Communication (VLC); and 
(iii) the {energy subsystem}, which captures both energy harvesting and consumption dynamics to enable autonomous and sustainable operation. 
At the core of our architecture lies the \texttt{ns3::Node} class, which encapsulates modules governing the node's communication interfaces, mobility behavior, and energy dynamics. This digital abstraction enables realistic and flexible simulation of hybrid optical--radio IoT environments. 

As detailed in Figure~\ref{fig:architecture}, the architecture follows a modular design to support extensibility and cross-layer interaction, as follows.  
\begin{itemize}
    \item Communication Interfaces: Both {BLE} and {OWC} interfaces are implemented as \texttt{NetDevice} subclasses, enabling independent configuration and protocol operation within the same node.  
    \item {Mobility Modeling:} A custom module is implemented, i.e.,  \texttt{OrientationAwareMobilityModel}, to capture the spatial orientation and motion of nodes, allowing accurate simulation of {directional optical links} and dynamic connectivity. 
    \item {Energy Management:} The node's energy behavior is represented through multiple \texttt{DeviceEnergyModel} instances linked to a centralized \texttt{BasicEnergySource}, providing detailed monitoring of energy consumption and flow across components. 
    \item {Optimization Integration:} A flexible {placeholder module} is included to accommodate algorithmic components, such as optimization algorithms, allowing future extensions without altering the base node structure. 
\end{itemize}

The implemented DT of the {RIoT node} supports hybrid communication using both BLE and VLC technologies. 
Both interfaces can operate simultaneously, enabling {full-duplex} communication for uplink and downlink data exchange. The BLE interface provides bidirectional radio connectivity, while the optical interface utilizes visible light for downlink transmission and infrared for uplink, ensuring reliable and energy-efficient hybrid communications. The {BLE} and {OWC} models in {NS-3} are developed to closely mirror the behavior and configuration of real {RIoT} hardware components. Each model follows the modular design principles of {NS-3}, ensuring a clean separation of layers and facilitating future extensions. 

The developed DT architecture also supports future integration with real-world node telemetry and optimization feedback loops. This design enables seamless interaction between DT and physical environments, supporting hybrid experimentation and online algorithm validation through the following components: 
\begin{itemize}
    \item The \texttt{LocalOptimizationAlgorithms} module allows the implementation of adaptive control strategies driven by real-time telemetry, such as sensed energy levels, mobility patterns, or channel conditions. 
    \item The \texttt{MobilityModel} can be synchronized with external data sources, including trajectory datasets or live positioning feeds, to replicate realistic movement and environmental dynamics. 
\end{itemize} 
This closed-loop capability facilitates online testing and hardware-in-the-loop integration, where {NS-3} simulations and physical node deployments operate and exchange data in real time, accelerating the validation of energy-aware and reconfigurable IoT algorithms. 

Finally, the RIoT node DT is designed to be calibrated and validated using empirical data to ensure high fidelity with real-world performance. For instance, 
\begin{itemize}
    \item {Energy model parameters} are derived from real-world measurements of BLE and OWC hardware, utilizing power and energy data collected directly from the target platforms, as detailed in Section~\ref{sec:Modeling}; 
    \item BER/SNR models are validated against controlled laboratory measurements of both RF and optical links to ensure realistic channel behavior; 
    \item End-to-end communication cycles and energy consumption patterns are aligned with observed hardware behavior to replicate operational dynamics accurately. 
\end{itemize}
Through this calibration process, our simulation environment can behave as a faithful digital twin of the physical system, supporting advanced tasks such as network testing, planning, and optimization of multi-modal communication strategies.

\subsection{OWC Module }

The {OWC module} is developed from the ground up to address architectural limitations identified in previous implementations. It introduces several key features that enhance realism, modularity, and interoperability within the {NS-3} environment: 
\begin{itemize}
    \item {Structured Physical and MAC layers}, providing a clear separation of functionalities and supporting protocol extensibility. 
    \item {Directional optical channel modeling} based on physical light propagation principles, enabling accurate simulation of alignment-dependent links. 
    \item {Full integration with the {NS-3} energy framework}, allowing precise tracking of energy consumption and interaction between communication and power subsystems. 
\end{itemize}

The {OWC model} enables detailed configuration of {optical parameters}, including transmission power, LED semi-angle, and photodetector field-of-view. It simulates {bit error rate (BER)} and {signal-to-noise ratio (SNR)} using realistic physical models based on optical channel characteristics, providing a high-fidelity representation of light-based communication behavior \cite{ribeiro2025short}.  

The developed OWC module simulates bi-directional optical wireless communication by modeling separate spectral bands and distinct physical-layer characteristics for transmission and reception. It is designed for full-duplex communication. Downlink is handled via visible light (e.g., white LEDs), while uplink transmissions use infrared (IR). A basic TDMA scheme, used by VLC, manages medium access by assigning time slots for transmission and reception, preventing overlap and collisions.  

\subsection{BLE Communication Module}

The {BLE module} is adapted from a community-developed implementation and updated for compatibility with {NS-3.40}. It models the essential behavior of Bluetooth Low Energy communication in RIoT nodes while maintaining modularity, energy traceability, and interoperability with other subsystems. The module features a stable and modular MAC/PHY structure and supports traffic generation at the MAC layer. To reduce complexity and focus on periodic data exchange, nonessential protocol elements such as device discovery and connection procedures are excluded, and all connections are statically configured, making this approach well-suited for star-topology, low-traffic IoT scenarios. 

Two PHY modes are implemented: 1~Mbit/s and 2~Mbit/s, which affect both bandwidth and energy consumption. While the 2~Mbit/s mode enables higher throughput, it requires a higher SNR and consequently reduces the effective communication range. Although empirical range measurements are not automated, our simulator employs standard propagation models (e.g., Friis) that, together with PHY rate selection, reproduce the expected shorter link ranges at 2~Mbit/s due to reduced link margin. Each BLE node can be individually configured to operate at either PHY rate, allowing the simulation of mixed-speed networks and adaptive rate strategies.  

\begin{figure}[t!]
	\centering
		\scalebox{1.25}{\includegraphics[width=0.34 \textwidth]{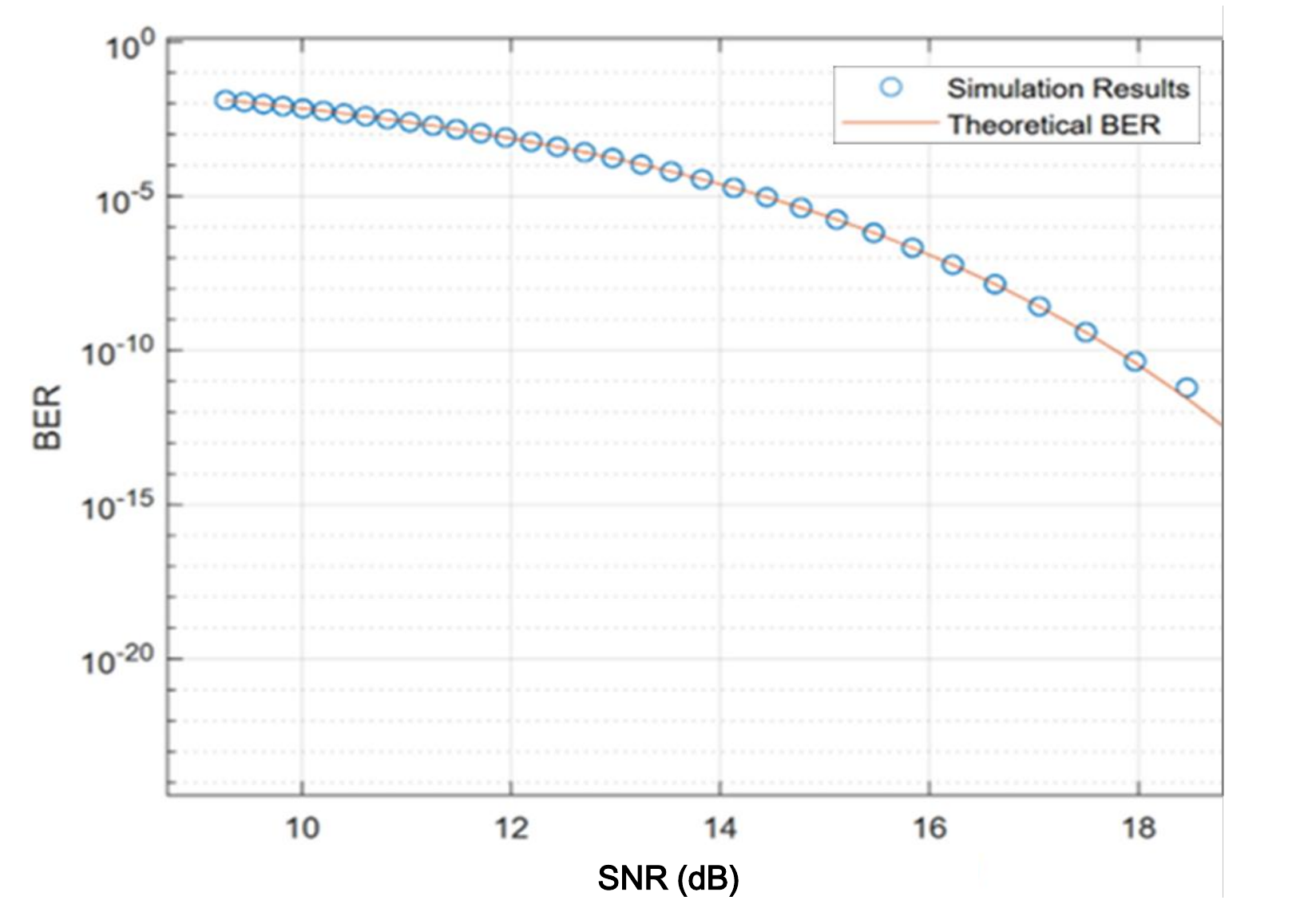} } 
	\caption{ Validation of the implemented GFSK modulation in NS-3 through BER versus SNR comparison with MATLAB reference results.   } 
	\label{fig:GFSK}
\end{figure}

Gaussian Frequency Shift Keying (GFSK), the modulation used in BLE, is implemented and validated. 
The comparison between the theoretical and implemented GFSK results is shown in Figure \ref{fig:GFSK}.  The BER--SNR curve obtained from the implemented GFSK model in {NS-3} closely matches the MATLAB reference results, exhibiting identical behavior up to an SNR of 18~dB. This agreement validates the correctness of the modulation and demodulation processes within the practical operating range of typical IoT devices, which generally operate under moderate SNR conditions due to short communication distances and low transmit power. Beyond this range, deviations have minimal impact on performance assessment, as IoT links rarely achieve or require higher SNR values. Hence, the observed correspondence up to 18~dB confirms that the proposed implementation provides a sufficiently accurate and reliable representation of GFSK performance for low-power IoT scenarios. 

\subsection{Integration of OWC and BLE Modules} 

\begin{figure}[t!]
	\centering
		\scalebox{1.45}{\includegraphics[width=0.34 \textwidth]{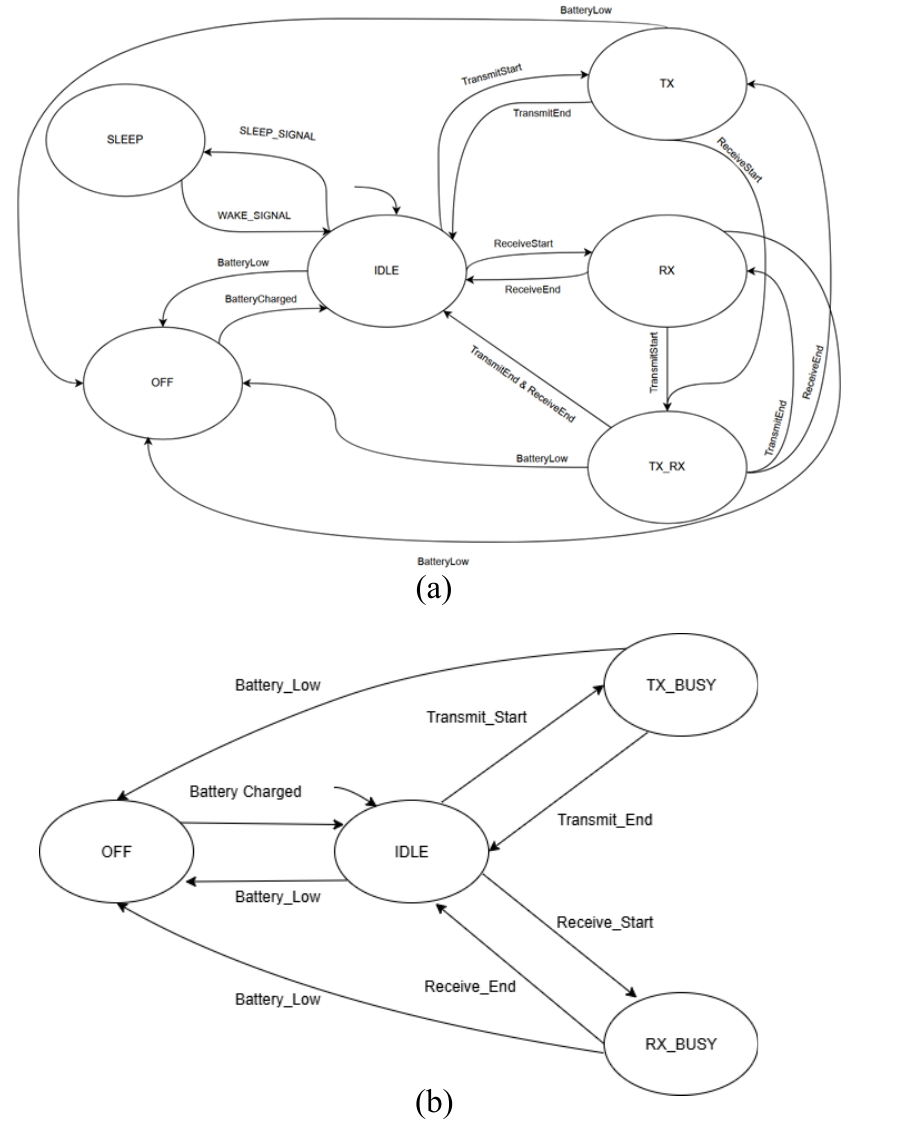} } 
	\caption{ Finite State Machine representation of the communication interfaces: (a) OWC, and (b) BLE.    } 
	\label{fig:FSM}
\end{figure}

Each communication interface is modeled using a finite state machine (FSM) that governs its operational state, i.e., Transmit, Receive, Idle, or Off, at any given time. Transitions between states dynamically update the node's energy consumption and propagate corresponding events to the mobility and application layers.  
The FSM of OWC, illustrated in Figure \ref{fig:FSM}-(a), comprises six main states: OFF, SLEEP, IDLE, TX (transmitting), RX (receiving), and TX-RX (simultaneous transmit and receive). Transitions between these states are triggered by specific events, such as TransmitStart, ReceiveStart, TransmitEnd, ReceiveEnd, SLEEP-SIGNAL, WAKE-SIGNAL, and battery-related events (e.g., BatteryLow, BatteryCharged). This event-driven structure enables realistic modeling of communication behavior and energy dynamics across protocol layers.  
Similarly, the FSM of BLE is presented in Figure \ref{fig:FSM}-(b), with the main states being IDLE, OFF TX-BUSY and RX-BUSY, controlled by a very similar set of triggers and events to the OWC state machine. 

To efficiently manage communications with energy-constrained RIoT nodes, we implement a polling-based medium access control (MAC) protocol that optimizes the operation of both the OWC and BLE modules. This protocol facilitates centralized coordination of node activity, ensuring high energy efficiency and scalability within the hybrid BLE/OWC communication environment. 
By synchronizing the transmission and reception cycles of individual nodes, the network achieves greater reliability, reduced collisions, and improved overall energy performance. The polling-based design further minimizes unnecessary power consumption by precisely scheduling node activity and avoiding idle listening or contention periods that typically waste energy in distributed access schemes. 
In this setup, the network controller periodically polls each node, triggering its transition from a low-power sleep state to an active state for data exchange. When not polled, nodes remain in an ultra-low-power or dormant mode, thereby significantly conserving energy and extending their operational lifetime. This controlled polling mechanism ensures deterministic access, predictable latency, and efficient energy utilization across the network.


\subsection{Energy Consumption and Harvesting Modules}

Energy autonomy is a defining characteristic of the RIoT node. By considering printed electronics and ambient energy harvesting (e.g., solar), the nodes operate without batteries or external power sources, storing energy in supercapacitors \cite{capuzzo2021ns}. Consequently, accurate energy modeling is a core requirement of the DT, making energy awareness an integral aspect of the RIoT node simulation.  
Thus, in our DT, the developed energy model comprises three core components that collectively capture the processes of energy storage, harvesting, and consumption:  
\begin{enumerate}
    \item Energy Buffer (source): A \texttt{BasicEnergySource} models the node's energy storage element (e.g., battery or capacitor), with a predefined maximum capacity expressed in joules.      
    \item Energy Harvesting: Our RIoT node supports solar energy harvesting through a photovoltaic harvester model, which periodically injects energy into the \texttt{BasicEnergySource}. The \texttt{EnergyHarvester} module simulates ambient charging by applying configurable input power profiles that emulate varying environmental light conditions. 
    \item Energy Consumption: Custom \texttt{DeviceEnergyModel} classes for both BLE and OWC interfaces define the instantaneous energy consumption in different operational states. Specifically, two energy consumption models have been implemented, i.e., constant current model, assuming fixed power draw per device state, and linear model, where current draw dynamically varies with transmission power, baud rate, and packet size.    
\end{enumerate} 
These modules enable real-time tracking of node energy dynamics and support realistic behaviors such as energy depletion, charging cycles, and adaptive operation based on available energy levels. 

\begin{figure}[t!]
	\centering
		\scalebox{1.45}{\includegraphics[width=0.34 \textwidth]{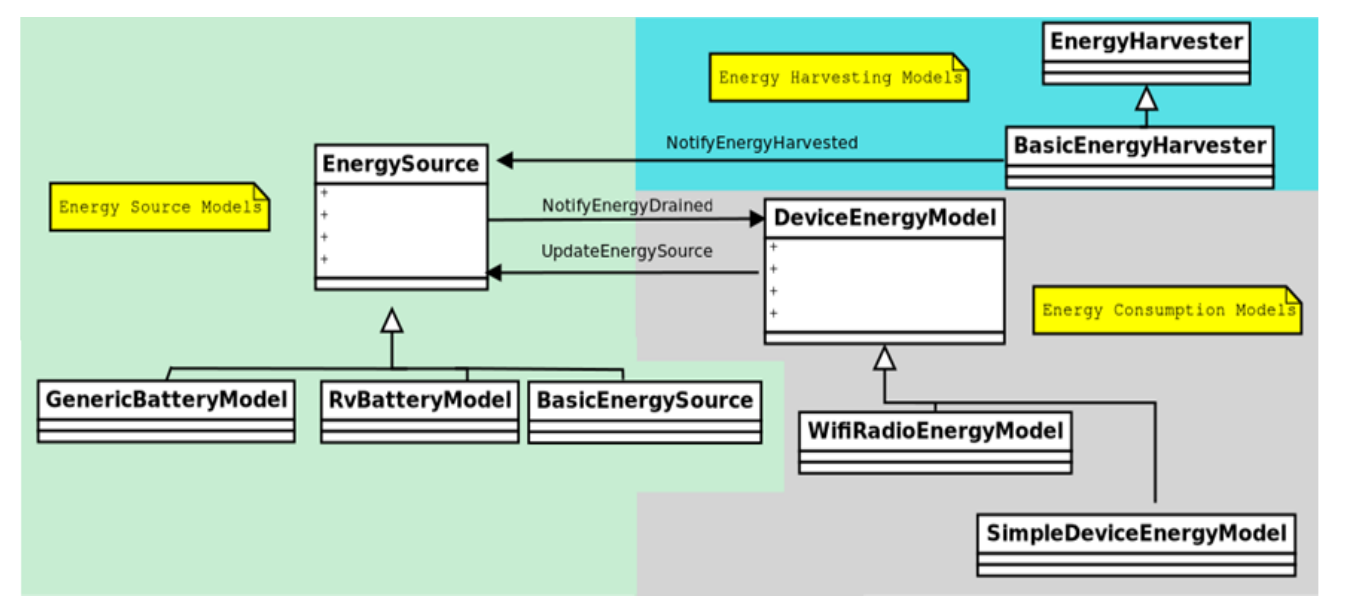} } 
	\caption{ Overview of the energy framework developed in NS-3.  } 
	\label{fig:energy}
\end{figure}

To accurately capture the heterogeneous energy usage of hybrid optical--radio IoT systems, custom \texttt{DeviceEnergyModel} classes were developed for the BLE and OWC modules, as well as for peripheral components such as screen updates and localization functions (see Figure~\ref{fig:energy}).  
The proposed models introduce fine-grained, state-dependent power characterization that reflects distinct operational states, including transmission, reception, idle, and sleep. Each model supports dynamic reconfiguration, allowing energy consumption parameters to adapt in real time based on transmission power, data rate, and node energy status. This modular and extensible design not only enables precise energy profiling but also establishes the foundation for cross-layer optimization and energy-aware control, where configuration parameters can be proactively adjusted to maximize energy efficiency and RIoT node lifetime. 
Furthermore, the energy models are calibrated using real-world measurements from the implemented RIoT hardware to accurately reflect actual energy consumption. The energy measurement setup used for this calibration is described in the following section.   



\section{ Energy Consumption Measurements \label{sec:Modeling} } 

In this section, we describe the energy measurement setup used to calibrate our DT models, detailing the procedures for capturing node-level energy consumption of RIoT nodes as well as the energy consumption of hybrid RF/OWC access points. 

\subsection{Energy Measurement Setup}

\begin{figure}[t!]
\centering
\subfloat[]{\includegraphics[width=0.75\linewidth]{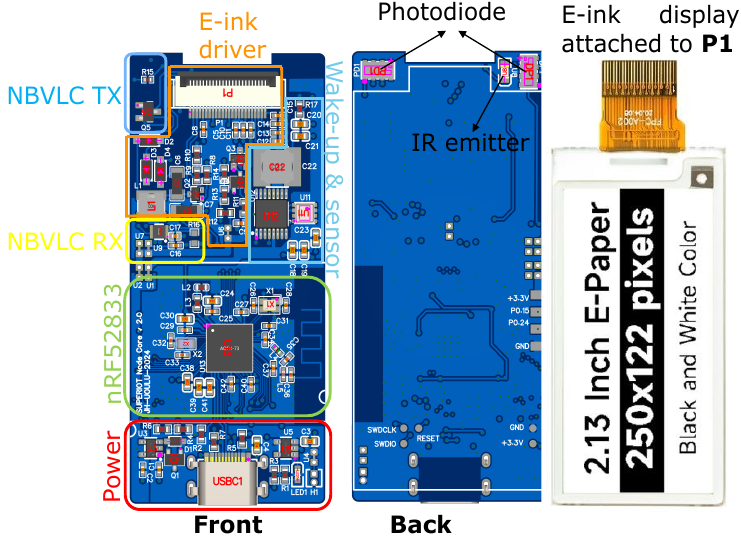}\label{fig:node}}
\hfil
\subfloat[]{\includegraphics[width=0.65\linewidth]{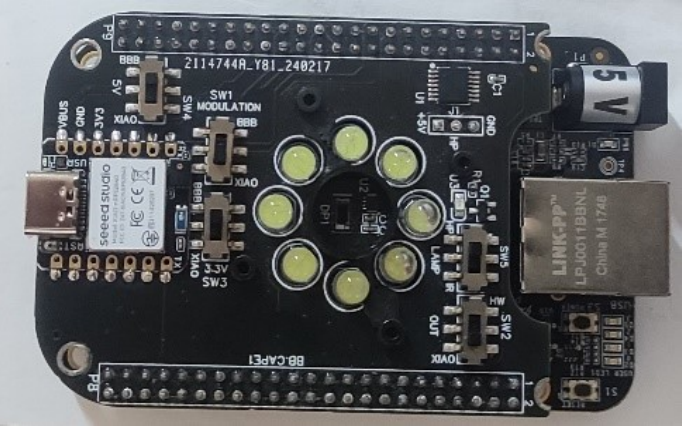}\label{fig:bbbap}}
\caption{Illustration of (a) Custom-engineered, multi-functional Si-based RIoT node, and (b) BBB Access Point (BBB platform + mini-lamp VLC gateway mounted on a Cape).}
\end{figure}

In our experiments, the silicon-based RIoT node illustrated in Figure~\ref{fig:node} has been used. This node integrates key hardware components for hybrid wireless communication and low-power operation. 
At its core, the nRF52833 Bluetooth Low Energy (BLE) system-on-chip (SoC) serves as the primary controller, managing sensing, communication, and peripheral interfaces. The node also includes a VLC transceiver, enabling optical data transmission and reception to complement BLE connectivity. Environmental monitoring is achieved using the BME688 sensor, which measures temperature, humidity, pressure, and gas concentration with high accuracy. Power-efficient wake-up functionality is provided by the AS3933-BTST, a low-power, light-based wake-up and timing IC that enables transitions from deep sleep to active mode in response to optical triggers. For visual output, the node employs a 2.13-inch monochrome E-ink display (250 × 122~pixels) with a full refresh time of approximately 2~s, offering a low-power solution for persistent visual feedback.

\begin{figure}[t]
    \centering
    \includegraphics[width=1\linewidth]{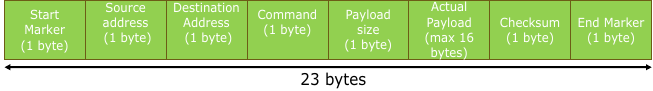}
    \caption{VLC frame structure for transmitting and receiving data/commands between node and AP.}
    \label{fig:vlc_frame}
\end{figure}

\begin{figure}[t]
    \centering
    \includegraphics[width=1\linewidth]{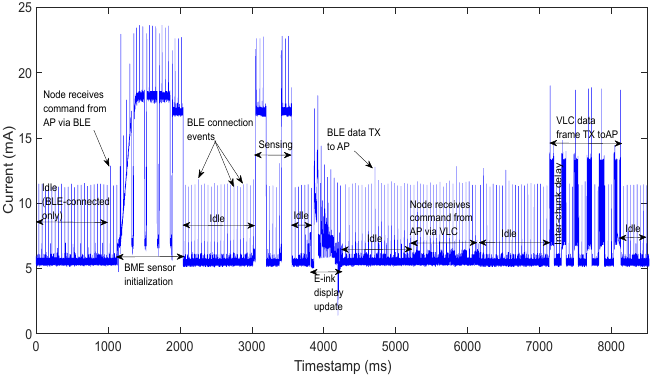}
    \caption{Current profile during different RIoT node operations.}
    \label{fig:normal_profile}
\end{figure}

Environmental sensor data and, optionally, localization coordinates—collectively referred to as the payload—are transmitted using a custom VLC frame structure consisting of 23~bytes, segmented into multiple 32-bit chunks to comply with the NEC protocol’s 32-bit limitation. As illustrated in Fig.~\ref{fig:vlc_frame}, each frame includes essential fields such as source and destination addresses, synchronization markers (start and end), payload type and size indicators, the actual payload, and a checksum for data integrity verification. In the current profile shown in Fig.~\ref{fig:normal_profile}, the VLC downlink and uplink transmissions appear as six distinct bursts, each corresponding to a 32-bit frame chunk. To ensure robust decoding at the receiver, an inter-chunk delay ($T_\text{d}$) of 100~ms is applied, allowing sufficient processing time for frame validation and payload extraction. 

The main access point (AP), shown in Fig.~\ref{fig:bbbap}, is built on the BeagleBone Black (BBB) platform and equipped with a custom-designed VLC lamp implemented on a BBB Cape. This lamp interfaces with the RIoT node to support bidirectional command and data exchange via both VLC and BLE. The system integrates a BLE-capable Seeed Studio Arduino nRF52840 module, along with dedicated photodiodes and LEDs to realize VLC functionality. The BBB AP can be connected to broader network infrastructure—comprising routers, switches, and both local (e.g., Raspberry~Pi-based) and cloud-hosted MQTT brokers—enabling remote bidirectional communication, data synchronization, and management across the IoT ecosystem.

Comprehensive energy characterization of the RIoT node and BBB AP was conducted using the Nordic Semiconductor Power Profiler Kit~II (PPKII). The PPKII operated in power supply mode, providing a regulated 3.3~V output to the RIoT node while concurrently recording current consumption across all operating phases. The BBB AP was powered at 5~V under equivalent monitoring conditions. Each PPKII was connected to a laptop via USB, and current measurements were acquired and logged using the \textit{nRF Connect for Desktop} software suite from Nordic Semiconductor.

\subsection{Node-level Energy Consumption}  
This subsection presents experimental measurements of RIoT node energy consumption under different communication and operating modes. 

When powered on, the RIoT node initiates BLE advertising in fast mode for 30~s (timeout period). If no connection is established within this period, it transitions to a slow advertising mode with an interval of 152.5~ms. After establishing a BLE connection with the AP, the node’s current consumption profile under various operational states is shown in Fig.~\ref{fig:normal_profile}. The measurement captures dynamic transitions across multiple functional stages, including BME688 sensor initialization, environmental sensing, E-ink display updates, VLC data transmission (TX) and command reception (RX), as well as BLE data transmission and command reception. These activities are interleaved with idle phases while the node remains connected to the AP via BLE. In this configuration, the BLE connection interval is set to 45~ms, with a transmission power of 0~dBm. It is worth noting that the BME688 sensor requires initialization only once—typically upon power-up—unless the node subsequently enters a deep-sleep state, in which case re-initialization is necessary to ensure accurate sensor readings. 
The initialization phase of the BME sensor, illustrated in Fig.~\ref{fig:normal_profile}, is relatively energy intensive, lasting approximately 916~ms and drawing an average current of about 14-15~mA.

\begin{table}[t]
\centering
\caption{Current draw measured across the node for different TX power levels during the BLE advertising state ($T_\text{advInt}$ = 152.5 ms).}
\label{tab:ble_adv_txpower}
\resizebox{\columnwidth}{!}{%
\begin{tabular}{c|ccc}
\toprule
\multirow{2}{*}{\textbf{$P_\text{tx}$ (dBm)}} &
\textbf{$I_\text{advInt}$ (mA)} &
\shortstack{\textbf{$I_\text{advIdle}$ (mA)} \\ $T_\text{advIdle}$ = 148.32 ms} &
\shortstack{\textbf{$I_\text{advEvent}$ (mA)} \\ $T_\text{advEvent}$ = 4.18 ms} \\
\cmidrule(lr){2-4}
& \multicolumn{3}{c}{\textit{Normal / Low-power}} \\
\midrule
+8 & 5.75 / 1.59 & 5.48 / 1.32 & 15.78 / 11.62 \\
+4 & 5.66 / 1.52 & 5.48 / 1.32 & 12.58 / 8.62 \\
0  & 5.58 / 1.45 & 5.46 / 1.32 & 9.65 / 5.95 \\
\bottomrule
\end{tabular}
}
\end{table}

\begin{table}[t]
\centering
\caption{Current draw measured across the node for different BLE TX power levels during the normal and low-power BLE connected-only state ($T_\text{connInt}$ = 45 ms).}
\label{tab:ble_conn_txpower}
\resizebox{\columnwidth}{!}{%
\begin{tabular}{c|ccc}
\toprule
\multirow{2}{*}{\textbf{$P_\text{tx}$ (dBm)}} &
\textbf{$I_\text{connInt}$ (mA)} &
\shortstack{\textbf{$I_\text{connIdle}$ (mA)} \\ $T_\text{connIdle}$ = 42.86 ms} &
\shortstack{\textbf{$I_\text{connEvent}$ (mA)} \\ $T_\text{connEvent}$ = 2.14 ms} \\
\cmidrule(lr){2-4}
& \multicolumn{3}{c}{\textit{Normal / Low-power}} \\
\midrule
+8 & 5.59 / 1.48 & 5.44 / 1.33 & 8.58 / 4.69 \\
+4 & 5.56 / 1.46 & 5.43 / 1.33 & 8.12 / 4.20 \\
0  & 5.51 / 1.43 & 5.43 / 1.33 & 7.31 / 3.52 \\
\bottomrule
\end{tabular}
}
\end{table}

\begin{table}[t]
\centering
\caption{Measured current consumption across the node for various BLE connection intervals and transmission power levels, comparing normal and low-power modes during a full operational cycle that includes sensing, optimized E-ink display, and associated idle periods.}
\label{tab:normal_lp_firmware01}
\resizebox{\columnwidth}{!}{%
\begin{tabular}{cc|cccc}
\toprule
\multicolumn{2}{c|}{} & \multicolumn{4}{c}{\textbf{Average current consumption (mA)}} \\
\multicolumn{2}{c|}{} & \multicolumn{4}{c}{\textit{Normal / Low-power}} \\
\midrule
\textbf{$T_\text{connInt}$} & \textbf{$P_\text{tx}$} & 
\shortstack{\textbf{$I_\text{sens}$} \\ $T_\text{sens}$ = 516 ms} & 
\shortstack{\textbf{$I_\text{idleSens}$} } &
\shortstack{\textbf{$I_\text{eink}$} \\ $T_\text{eink}$ = 0.435 s} &
\shortstack{\textbf{$I_\text{idleEink}$}} \\ 
\midrule
\multirow{3}{*}{11.25} & +8 & 12.75 / 8.76 & 5.95 / 1.92 & 7.69 / 4.12 & 6.03 / 2.12 \\
 & +4 & 12.65 / 8.64 & 5.89 / 1.85 & 7.58 / 4.03 & 5.91 / 2.04 \\
 & 0  & 12.55 / 8.52 & 5.79 / 1.71 & 7.45 / 3.93 & 5.84 / 1.90 \\
\midrule
\multirow{3}{*}{45 (default)} & +8 & 12.30 / 8.28 & 5.58 / 1.49 & 7.28 / 3.66 & 5.62 / 1.68 \\
 & +4 & 12.28 / 8.26 & 5.56 / 1.46 & 7.27 / 3.65 & 5.59 / 1.64 \\
 & 0  & 12.26 / 8.23 & 5.50 / 1.43 & 7.24 / 3.63 & 5.56 / 1.63 \\
\midrule
\multirow{3}{*}{250} & +8 & 12.20 / 8.17 & 5.47 / 1.37 & 7.20 / 3.53 & 5.53 / 1.55 \\
 & +4 & 12.18 / 8.15 & 5.47 / 1.36 & 7.20 / 3.52 & 5.53 / 1.54 \\
 & 0  & 12.17 / 8.14 & 5.47 / 1.35 & 7.19 / 3.52 & 5.50 / 1.54 \\
\midrule
\multirow{3}{*}{1000} & +8 & 12.13 / 8.13 & 5.45 / 1.35 & 7.19 / 3.50 & 5.50 / 1.53 \\
 & +4 & 12.12 / 8.13 & 5.45 / 1.33 & 7.17 / 3.50 & 5.50 / 1.53 \\
 & 0  & 12.09 / 8.13 & 5.45 / 1.33 & 7.17 / 3.50 & 5.49 / 1.53 \\
\bottomrule
\end{tabular}
}
\end{table}

In practical deployments, the RIoT node communicates with the access point (AP) using both BLE and VLC. Under normal operation, it supports hybrid connectivity, enabling bidirectional (uplink and downlink) communication over both technologies. 
As illustrated in Fig.~\ref{fig:normal_profile}, the AP can issue downlink commands via either BLE or VLC, prompting the node to transmit corresponding uplink data such as sensor readings. 

Table \ref{tab:ble_adv_txpower} shows the measured average current consumption of the BLE advertising interval, $I_{\text{advInt}}$, for different TX power levels ($P_\text{tx}$).
During BLE advertising events when the device is actively broadcasting its presence, higher transmit power levels result in higher average current consumption for the advertising event, $I_{\text{advEvent}}$. 
Idle current consumption between advertising events, $I_{\text{advIdle}}$, remains relatively constant across different TX power levels, as expected.
Similarly, Table \ref{tab:ble_conn_txpower} presents the average current consumption of the node during BLE connection intervals for different transmission power levels ($P_\text{tx}$). A clear reduction in current is observed as the BLE TX power decreases from +8~dBm to 0~dBm, indicating that lower transmission power substantially reduces the current drawn during active BLE connection events. In contrast, the idle current $I_{\text{connIdle}}$, measured between connection events, remains nearly constant, confirming that idle periods are unaffected by variations in TX power, as expected.

Table~\ref{tab:normal_lp_firmware01} summarizes the measured average current consumption over a complete operational cycle that includes sensing and E-ink display updates, along with their respective idle periods. The measurements are presented for different BLE connection intervals ($T_\text{connInt}$) and transmission power levels ($P_\text{tx}$). In the table, $T_{\text{sens}}$ denotes the average duration of a sensing operation, and $I_{\text{sens}}$ represents the corresponding average current. Similarly, $T_{\text{eink}}$ and $I_{\text{eink}}$ refer to the duration and average current of the E-ink display operation, respectively. The parameters $I_{\text{idleSens}}$ and $I_{\text{idleEink}}$ indicate the average current consumption during the idle periods following the sensing and E-ink display operations, respectively.

Figure~\ref{fig:normal_profile} illustrates the node transmitting data to the AP using the BLE UART service. The data is sent as a 116-byte JSON string containing the Source ID, Destination ID, and sensor measurements such as temperature, humidity, pressure, and gas concentrations, and dummy location coordinates (X, Y, Z). The BLE link operates with a 2 Mbps PHY data rate, a Maximum Transmission Unit (MTU) of 247 bytes, a connection interval ($T_\text{connInt}$) of 45 ms, and a transmission power ($P_\text{tx}$) of 0 dBm.
BLE uplink transmissions from the node to the AP are brief, lasting approximately 3.13 ms, with current draw ranging from 9.10 mA at 0 dBm to 13.41 mA at +8 dBm. Downlink transmissions from the AP to the node are around 2.33 ms, drawing between 7.36 mA at 0 dBm and 7.89 mA at +8 dBm. In downlink operations, the AP writes directly to the appropriate node's BLE characteristics service, enabling it to request sensor data or trigger actions, such as refreshing the E-ink display.

As illustrated in Fig.~\ref{fig:normal_profile}, the RIoT node is also capable of receiving commands in the downlink and transmitting the corresponding data in the uplink via VLC, following the frame structure depicted in Fig.~\ref{fig:vlc_frame}, while maintaining an active BLE connection with the AP. During VLC downlink operation (command reception from the AP), the node’s current consumption remains comparable to the BLE-connected idle level (around 5.84 mA for one VLC frame chunk), whereas uplink VLC transmission incurs a noticeably higher current draw (around 9.15 mA for one VLC frame chunk).

Assuming a BLE connection interval of 45 ms, a transmission power of 0 dBm, and a supply voltage of 3.3 V, the energy required for an uplink BLE transmission is approximately 94 µJ, based on a measured current of 9.10 mA over 3.13 ms. In contrast, transmitting the same data via VLC in the uplink—segmented into six chunks while maintaining the BLE connection—requires roughly 21.5 mJ. This corresponds to an average current of 7.17 mA over a total duration of 0.91 s (with each VLC chunk averaging 9.15 mA over 68 ms). These results indicate that BLE is substantially more energy-efficient for uplink data transfer when a connection is already established, primarily due to its higher data rate and shorter transmission duration. Nonetheless, VLC remains a viable alternative in environments where BLE operation is restricted, such as in medical settings with sensitive equipment (e.g., X-ray imaging systems).

\subsection{Access Point Energy Consumption\label{sec:ap_energy} } 

Table~\ref{tab:bbb_power_tests} presents the measured average current and power consumption of the BBB–based AP illustrated in Fig.~\ref{fig:bbbap}, evaluated under various operating configurations.
For test IDs~1–5, the VLC LEDs on the BBB CAPE were disabled (OFF), and the BLE interface on the Seeed Xiao nRF52840 module remained inactive.
During the initial boot sequence (test~ID~1), the AP draws an average current of approximately 405~mA (2.03~W). In this configuration, the BBB's USB~2.0 port is connected to the Type-C interface of the Xiao nRF52840 module on the lamp CAPE, while the Ethernet interface provides a wired connection to a network switch. The boot process lasts about 72~s, after which the system transitions to an idle state. In the idle state (test~ID~2), with both USB and Ethernet links active, the current draw decreases to 255~mA (1.28~W).
Additional idle measurements were conducted with only the Ethernet cable connected (test~ID~4) and with both USB and Ethernet disconnected (test~ID~3). A comparison of these results indicates that enabling the USB interface between the BBB and the Xiao module increases the average current by approximately 14~mA (tests~2 vs.~4), while activating the Ethernet interface adds roughly 71~mA (tests~3 vs.~4), even in the absence of active data transfer.
\begin{table}[t]
\caption{Measured current and power of BBB Access Point under different conditions.}
\label{tab:bbb_power_tests}
\centering
\footnotesize 
\setlength{\tabcolsep}{2.5pt} 
\begin{tabular}{c l r r}
\toprule
\textbf{ID} & \textbf{Condition} & \textbf{I (mA)} & \textbf{P (W)} \\
\midrule
1 & Boot: USB+Eth (VLC/BLE OFF) & 405 & 2.03 \\
2 & Idle: USB+Eth (VLC/BLE OFF) & 255 & 1.28 \\
3 & Idle: No USB/Eth (VLC/BLE OFF) & 170 & 0.85 \\
4 & Idle: Eth only (VLC/BLE OFF) & 241 & 1.21 \\
5 & TX: USB+Eth (VLC/BLE OFF) & 388 & 1.94 \\
6 & TX: USB+Eth (VLC 98\%, BLE OFF) & 590 & 2.95 \\
\bottomrule
\end{tabular}
\end{table}

To assess active communication performance, iPerf3 was employed to generate bidirectional Ethernet traffic between the BBB AP (client) and a laptop (server) connected through the same switch. Under continuous data transfer (test~ID~5), the average current increases to 388~mA, representing a rise of approximately 133~mA relative to the idle condition in test~ID~2.
When the VLC LEDs operate at 98\% brightness (test~ID~6), the maximum recorded power consumption of the BBB AP reaches approximately 2.95~W.

\begin{figure}
    \centering    \includegraphics[width=0.85\linewidth]{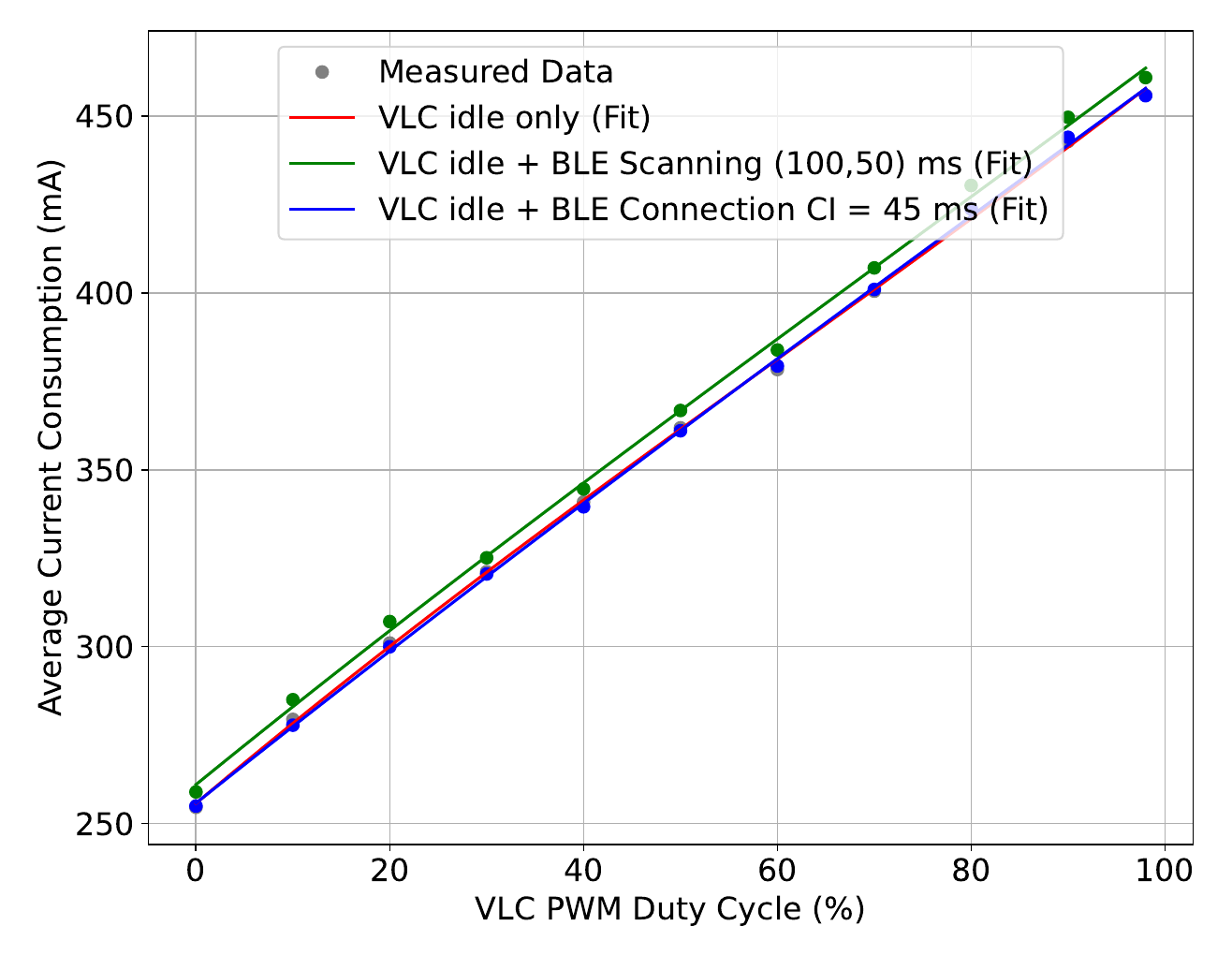}
    \caption{Average current consumption of BBB AP under three test cases. 
    }
    \label{fig:bbb_energy}
\end{figure}

To assess the influence of illumination intensity on system power consumption, the VLC PWM duty cycle—governing the brightness of the lamp LEDs—was varied, and its effect on the average current drawn by the BBB-based access AP was analyzed.
Figure~\ref{fig:bbb_energy} compares the measured average current under three operating conditions:
(1) VLC idle, with BLE disabled;
(2) VLC idle, with BLE scanning enabled (scan interval = 100~ms, scan window = 50~ms); and
(3) VLC idle, with an established BLE connection to the RIoT node (connection interval = 45~ms).
In all cases, the system remained idle, with no data exchange between the AP and the node. The BLE transmit power was fixed at 0~dBm, and both the USB and Ethernet interfaces on the BBB AP were connected but inactive (no active Ethernet traffic).
As shown in Fig.~\ref{fig:bbb_energy}, the average current exhibits an approximately linear increase with the VLC PWM duty cycle. When BLE scanning is enabled concurrently with VLC operation, the current consumption shows a nearly constant offset of about +5.3~mA relative to the VLC-only condition. In contrast, when a BLE connection is maintained without active data transmission, the current difference with respect to the VLC-only case remains negligible.

\section{Proactive Cross-Layer Optimization \label{sec:Optimization}}

In this section, we first define the key monitoring parameters within the DT framework used to track the performance and operational behavior of RIoT nodes. We then present a low-complexity optimization algorithm designed to configure these nodes adaptively. 
The objective is to develop a solution that is both effective and lightweight—suitable for implementation on low-capability hardware with minimal computational overhead, while maintaining an optimal balance between performance, energy efficiency, and operational feasibility in real-world deployments.  

\subsection{Monitoring Variables and Status Information}

First, a set of {core monitoring variables} is defined within the DT framework to track the performance, energy status, and operational behavior of RIoT nodes. These variables serve as input to the cross-layer optimization logic, allowing each node to adapt to network dynamics and proactively adjust its configuration to optimize energy consumption while maintaining QoS.  
The monitored variables include:  
\begin{itemize}
    \item {Signal-to-Noise Ratio (SNR):} Assesses channel quality and informs modulation, interface selection, and transmission decisions.  
    \item {Node Location:} Supports directional optimization (e.g., OWC beam alignment) and mobility-aware resource allocation.  
    \item {Packet Loss Ratio (PLR):} Reflects communication reliability and triggers fallback or retransmission strategies.  
    \item {Remaining Energy:} Expressed in voltage or Joules, determines node activity level and duty cycle.  
    \item {Transmission Power (TX Power):} Balances connectivity and energy efficiency as both a measurable and controllable parameter.  
    \item {Modulation and Coding Scheme (MCS):} Enables adaptive data rate and energy optimization.  
    \item {Active Interface(s):} Identifies whether BLE, OWC, or both are active, supporting multimodal management.  
    \item {Peripheral Activity (e.g., screen, localization):} Accounts for non-communication energy loads.  
\end{itemize}

Each variable contributes differently to performance and energy optimization, carrying a weight based on its system impact. For instance, high-priority variables (e.g., remaining energy and SNR) drive time-critical decisions such as modality switching. Medium- and low-priority variables (e.g., node location, peripheral activity) can influence longer-term adjustments such as spatial reuse or power budgeting.   This prioritization supports three objectives: (i) {QoS assurance}, maintaining reliable communication links,  (ii) energy efficiency, extending node lifetime, and  (iii) operational flexibility,  enabling rapid adaptation to dynamic network conditions.  Furthermore, the status information of RIoT nodes is gathered at different rates depending on variable criticality and node constraints. For instance, high-priority parameters are updated on a per-packet basis, while others follow configurable sampling intervals or are event-triggered when significant changes occur. This adaptive monitoring ensures accurate, real-time feedback to the optimization layer implemented in NS-3. 

The monitored variables, such as remaining energy, SNR, and node location, serve as real-time inputs to the optimization layer, enabling each node to anticipate network changes and adapt its configuration proactively. Leveraging these inputs, the optimization strategies jointly manages communication interfaces, energy states, and peripherals to balance performance, efficiency, and responsiveness.  
Furthermore, we define three distinct operational modes for the RIoT nodes: Performance Mode, Conservation Mode, and Sleep Mode. 
In Performance Mode, the node operates under normal conditions without restrictions on energy consumption or communication capabilities, ensuring maximum performance and responsiveness. 
In Conservation Mode, the node maintains reception functionality but limits its transmission rate to preserve energy. It also temporarily disables non-essential components such as screen updates and localization features. This mode is activated when the node's remaining energy is low but not yet critical, allowing continued operation while preventing premature battery depletion. 
In Sleep Mode, the node suspends all communication and sensing activities until a higher operational mode is reactivated. This mode is employed when the node's energy reaches a critical threshold or when communication is unnecessary for a defined period, thereby minimizing power usage and extending overall network lifetime.

   

\subsection{Energy-aware Utility-based RIoT Optimization} \label{subsec:EUNO}

For sustainable and energy-efficient operation of RIoT nodes, we propose the Energy-aware Utility-based Node Optimization (EUNO) algorithm. 
EUNO adopts a heuristic decision-making framework that enables each node to autonomously select the most appropriate operational action by evaluating a unified utility function. This function quantifies the expected benefit of each potential action based on the node’s current state, residual energy, traffic pattern, and network conditions. Each action is defined by its modality (either OWC or BLE) and operational mode (Performance, Conservation, or Sleep).   
The overall utility is formulated as a weighted combination of multiple sub-utilities, each representing a distinct operational function: communications modality, screen updates, localization, and predicted energy consumption. These sub-utilities guide the node to make decisions that achieve the best balance between performance and energy conservation.  
 
Below, we present the formulation of the overall utility function and its constituent sub-utilities.

\subsubsection{Overall Utility Function}
The overall utility associated with an action $a$ is defined as: 
\begin{equation}
U(a) = p_{{M}} \, U_{{M}}(a) 
+ p_{{S}} \, U_{{S}}(a)
+ p_{{L}} \, U_{{L}}(a)
+ p_{{E}}(t) \, U_{{E}}(a), 
\label{eq:tot}
\end{equation}
where $p_{{M}}$, $p_{{S}}$, and $p_{{L}}$ are static weights that represent the relative importance of each utility component according to the specific application requirements and traffic characteristics, such that $p_{{M}} + p_{{S}} + p_{{L}} = 1$.    
The energy-related weight $p_{{E}}(t)$ is time-dependent and dynamically adapts to the node's residual energy. 
It increases in significance as the energy level decreases, ensuring that energy preservation becomes a higher priority under heavy usage or low-energy conditions. Hence, it is defined as:   
\begin{equation}
p_{E}(t) = 1-\dfrac{f_r-f_{c}}{1-f_{c}}
\end{equation}
where $f_r$ is the normalized residual energy fraction and $f_{c}$ defines the critical energy threshold below which energy efficiency dominates the decision-making process. 

\subsubsection{Modality Utility}  
The modality utility, $U_{{M}}$, quantifies the trade-off between communication performance and energy efficiency based on the node's current residual energy. $U_{{M}}$ adapts the choice of communication modality according to the node's present operating mode and energy state.   
When operating in {Performance Mode}, the node prioritizes communication modalities offering higher data rates and stronger signal quality. Conversely, in {Conservation Mode}, it selects modalities that minimize energy consumption. To capture this adaptive behavior, the utility incorporates the remaining energy fraction $f_r$ and evaluates whether the next state corresponds to: 
(i) {Performance Mode}, weighted by $p_p$, favoring high throughput ($x_t$, weighted by $p_t$); or 
(ii) {Conservation Mode}, weighted by $p_c$, emphasizing energy efficiency ($x_e$, weighted by $p_e$). 
Furthermore, to prevent excessive switching between communication modalities (e.g., between OWC and BLE), a small penalty term $p_{ch}x_{ch}$ is introduced, where $p_{ch}$ is tuned based on historical data to discourage frequent mode oscillations.   
Formally, the modality utility is expressed as:
\begin{equation}
U_{{M}} = f_{r} \, (p_{p} x_{p} + p_{t} x_{t}) 
+ (1 - f_{r}) \, (p_{c} x_{c} + p_{e} x_{e}) 
- p_{ch} x_{ch},
\label{eq:Mod}
\end{equation}
This formulation allows the node to dynamically balance throughput and energy efficiency, favoring high-performance communication when energy is abundant and transitioning to low-power operation as the energy reserve decreases.

\subsubsection{Screen Utility} 
The screen utility, $U_{{S}}$, quantifies how well the node's screen activity aligns with user/application requirements. It rewards actions that activate the screen when user interaction is likely and penalizes unnecessary updates during idle periods. The utility is defined as:
\begin{equation}
U_S(a) = 
\begin{cases}
\alpha, & \text{if } \big(a.\text{state} = P \land \mathcal{R}(\text{s})\big) \\
        & \text{or } \big(a.\text{state} = C \land \lnot \mathcal{R}(\text{s})\big), \\
0, & \text{otherwise,}
\end{cases}
\label{eq:Screen}
\end{equation}
where $\alpha$ is the reward factor associated with actions that synchronize screen activity with usage needs. Here, $a.\text{state} = P$ indicates that the RIoT node is operating in {Performance Mode}, while $a.\text{state} = C$ corresponds to {Conservation Mode}.  
The function $\mathcal{R}(\text{s})$ determines whether a screen refresh is required based on the estimated probability of user interaction at the current time $t_{{n}}$:
\begin{equation}
    \mathcal{R}(\text{s}) = 
    \begin{cases}
        1, & \text{if } p_{\text{int}}(t_{{n}}) > \theta_s, \\
        0, & \text{otherwise,}
    \end{cases}
\end{equation}
where $p_{\text{int}}(t_{{n}})$ represents the likelihood of user interaction derived from recent traffic and usage statistics, and $\theta_s$ is a tunable decision threshold.  This formulation allows the node to adaptively manage screen activity according to both operational mode and user behavior patterns. 


\subsubsection{Localization Utility}  
The localization utility, $U_{{L}}$, evaluates how well the selected action aligns the node's localization activity with application demands and network dynamics. It rewards actions that activate localization when it is required, either by the application or due to mobility-induced changes in network topology, denoted by $\mathcal{R}(\text{l})$:
\begin{equation}
U_L(a) = 
\begin{cases}
\beta, & \text{if } \big(a.\text{state} = P \land \mathcal{R}(\text{l})\big) \\
       & \text{or } \big(a.\text{state} = C \land \lnot \mathcal{R}(\text{l})\big), \\
0, & \text{otherwise,}
\end{cases}
\label{eq:Loc}
\end{equation}
where $\beta$ is the reward factor associated with performing localization at appropriate times. 
The requirement condition $\mathcal{R}(\text{l})$ determines whether localization should be triggered. It depends on the node's mobility and the stability of its network links: 
\begin{equation}
    \mathcal{R}(\text{l}) =
    \begin{cases}
        1, & \text{if } p_{\text{m}}(t_{{n}}) > \theta_{{l}}, \\
        0, & \text{otherwise,}
    \end{cases}
\end{equation}
where $p_{\text{m}}(t_{{n}})$ represents the estimated probability of node mobility (or network conditions change) at time $t_{{n}}$, and $\theta_{{l}}$ is the corresponding decision threshold.    
This formulation allows the node to adapt its localization activity dynamically according to real-time network behavior and mobility patterns, ensuring efficient use of energy and communication resources.

To obtain $p_{\text{m}}(t_{n})$, an Exponentially Weighted Moving Average (EWMA)-based network predictor is implemented. This predictor uses the EWMA of the received SNR ($S(t_{n})$) to establish a stable baseline average, where 
\begin{equation}
    W_A(t_n) = \lambda \cdot S(t_{n}) + (1 - \lambda) \cdot W_A(t_n-1). 
\end{equation} 
Then, $p_{\text{m}}(t_n)$ is derived by mapping the deviation between the smoothed average and the current instantaneous SNR, i.e., $ \Delta S(t_n) = \big| W_A(t_n) - S(t_n) \big|$, onto a Sigmoid function: 
\begin{equation}
    p_{\text{m}}(t_n) = \frac{1}{1 + e^{-k \cdot (\Delta S(t_{n}) - C)}}. 
\end{equation}  
Here, $\lambda$ is the smoothing constant, and the Sigmoid parameters $k$ (slope) and $C$ (critical deviation threshold) are tuned to control the sensitivity and minimize false detections caused by short-term fading.  

\subsubsection{Energy Utility} 
The predicted energy utility, $U_{{E}}(a)$, evaluates the energy efficiency of action $a$ with respect to the node's residual energy and operational context. It rewards actions that minimize power consumption and penalizes those that accelerate depletion. It is defined as:
\begin{equation}
U_E(a) = 1 - \frac{E_a(t)}{E_{\max}},
\label{eq:UE}
\end{equation}
where $E_a(t)$ represents the predicted energy consumed by executing action $a$ during the current cycle, and $E_{\max}$ denotes the maximum possible energy that the node can accumulate. 


We define each RIoT node's action as a tuple $a = (\mathcal{S}, \mathcal{M})$, where $\mathcal{S} \in \{P, C, S\}$ denotes the next operating state ({Performance}, {Conservation}, or {Sleep}), and $\mathcal{M} \in \{\text{OWC}, \text{BLE}\}$ represents the selected communication modality. During each optimization cycle, the node evaluates all feasible actions $a \in \mathcal{A}$ according to their total utility $U(a)$, and selects the one yielding the highest value:
\begin{equation}
a^* = \arg\max_{a \in \mathcal{A}} U(a).
\end{equation}
This adaptive process enables each RIoT node to make context-aware decisions that dynamically balance energy consumption, communication performance, and application responsiveness. To prevent critical energy depletion, EUNO restricts all non-sleep actions when the remaining energy fraction $f_r$ of any RIoT node falls below a predefined threshold $f_{c}$. 
The main steps of the EUNO algorithm are summarized in Algorithm~\ref{alg:EUNO}.
\begin{algorithm}[b]
\caption{Energy-aware Utility-based Node Optimization (EUNO) Algorithm}
\label{alg:EUNO}
\begin{algorithmic}[1]
\Require Node state parameters $\mathcal{S}$, application requirements $\mathcal{R}$, remaining energy fraction $f_r$, weighting coefficients $\{p_{\text{M}}, p_{\text{S}}, p_{\text{L}}, p_{\text{E}}\}$
    \State Initialize action set $\mathcal{A} = \{a_1, a_2, \ldots, a_N\}$
    \If{$f_{r} < f_{c}$}
        \State Restrict all non-sleep actions and enter \emph{Sleep Mode}.
        \State \textbf{return}
    \EndIf
    \For{each action $a \in \mathcal{A}$}
        \State Estimate energy consumption $E_a(t)$ for the next interval. 
        \State Compute {Modality Utility} $U_{{M}}(a)$ using (\ref{eq:Mod}).
        \State Compute {Screen Utility} $U_{{S}}(a)$ using (\ref{eq:Screen}).
        \State Compute {Localization Utility} $U_{{L}}(a)$ using (\ref{eq:Loc}).
        \State Compute {Energy Utility} $U_{{E}}(a)$ using (\ref{eq:UE}). 
        \State Calculate total utility $U(a)$ using (\ref{eq:tot}). 
    \EndFor
    \State Select optimal action $a^* = \arg\max_{a \in \mathcal{A}} U(a)$. 
    \State Execute $a^*$ and update node configuration. 
\end{algorithmic}
\end{algorithm} 
We also remark that the proposed EUNO algorithm can be readily extended to an online learning paradigm \cite{11142858}, in which each node continuously evaluates and updates its communication and operational decisions based on real-time feedback, residual energy, and network conditions. This enables adaptive and self-optimizing behavior without the need for offline training or predefined policies. Observations such as energy depletion rate, link quality variations, and switching frequency can be exploited to iteratively adjust the decision parameters, including the utility weights ($p_p$, $p_t$, $p_c$, $p_e$, $p_{ch}$) and thresholds ($\theta_s$, $\theta_l$, $f_c$), allowing the system to progressively refine its decision-making policy over time.

\section{Performance Evaluation \label{sec:Evaluation}}

In this section, we first describe the simulation environment and parameter settings, followed by performance results evaluating the RIoT nodes, the developed DT framework, and the proposed optimization strategy. 

\subsection{Environment Setup \label{sec:Setup}}

Building on the developed DT described in Section~\ref{sec:DT_NS3}, a reconfigurable simulation environment was implemented using NS-3. The simulated network consists of multiple nodes arranged in a star topology, where several gateways are each connected to a set of RIoT nodes. Each RIoT node operates with limited energy resources and communicates exclusively with its assigned gateway.  

To emulate diverse traffic conditions and usage patterns, each node executes a custom application developed specifically for this evaluation. The application periodically transmits packets between node pairs at a fixed data rate, utilizing either OWC or BLE as the communication modality. Moreover, each RIoT node is equipped with an e-ink display and a localization module, both of which are considered in the simulation to accurately capture their impact on energy consumption and system performance. Each RIoT node is also equipped with an energy harvester that replenishes its energy at a fixed harvesting rate, enabling the node to recover energy during sleep periods. To ensure realistic performance evaluation, parameters such as signal-to-noise ratio (SNR), packet loss, and link quality are accurately modeled within the NS-3 framework. These values are dynamically computed based on factors such as node position, inter-node distance, antenna orientation, and other relevant physical parameters, thereby providing a comprehensive and realistic emulation of the network environment. 
The main simulation parameters used for performance evaluation are summarized in Table~\ref{tab:parameters}, and all simulations were conducted using the NS-3 network simulator. The static weights in the overall utility function were normalized to $1$ to maintain balanced, easy to understand overview. However, $U_{{M}}$ is within a $[-0.1, 4]$ bound while the rest of the modalities are within $[0, 1]$. The $\Delta t_{pr}$ and the physical attributes, such as the end node to gateway distance, $d_{e \to g}$, and the incidence angle between the end nodes and the gateway, which is relevant for OWC communication, were chosen with the intention of simulating a normal communication favorable scenario.

\begin{table}[t!]
\centering
\caption{Simulation Parameters.}
{  \footnotesize 
\begin{tabular}{|l|l|l|l|}
\hline
\textbf{Parameter}          & \textbf{Value}                & \textbf{Parameter}          & \textbf{Value}                \\ \hline
$p_{\text{M}}$      &             0.91               & $f_{{c}} $                   &           0.2              \\ \hline
$p_{{L}} $           &           0.045           & $p_{{S}}$                     &      0.045      \\ \hline
$p_{p}$                      &          2                  &            $p_{t}$          &       2        \\ \hline
$p_{c}$               &          1             &      $p_{e}$      &      0.8                 \\ \hline
$p_{ch}$         &              0.1             &          $\Delta t_{pr}$             &           10 s                \\ \hline
$d_{e \to g}$             &             1 m             &        $\angle_i$         &     30º      \\ \hline

\end{tabular}
} 
\label{tab:parameters}
\end{table}

\subsection{ Experimental Energy Consumption Evaluation \label{sec:Evaluation_Experimental}} 

To demonstrate the practical feasibility of the proposed framework, this subsection evaluates the real energy consumption of the RIoT node under different communication and operating modes. Given that the RIoT node is designed to operate without batteries, relying solely on ambient light and RF energy harvested and stored in printed supercapacitors, its energy availability is inherently limited and highly variable. While this enables a sustainable and eco-friendly architecture, it also makes energy-aware operation essential. Therefore, accurately characterizing the node's energy consumption is crucial to justify fine-grained optimization, calibrate the DT models, and ensure reliable real-world operation. 



\subsubsection{E-ink Display Optimization}
\begin{figure}[t]
\centering
\subfloat[]{\includegraphics[width=0.47\linewidth]{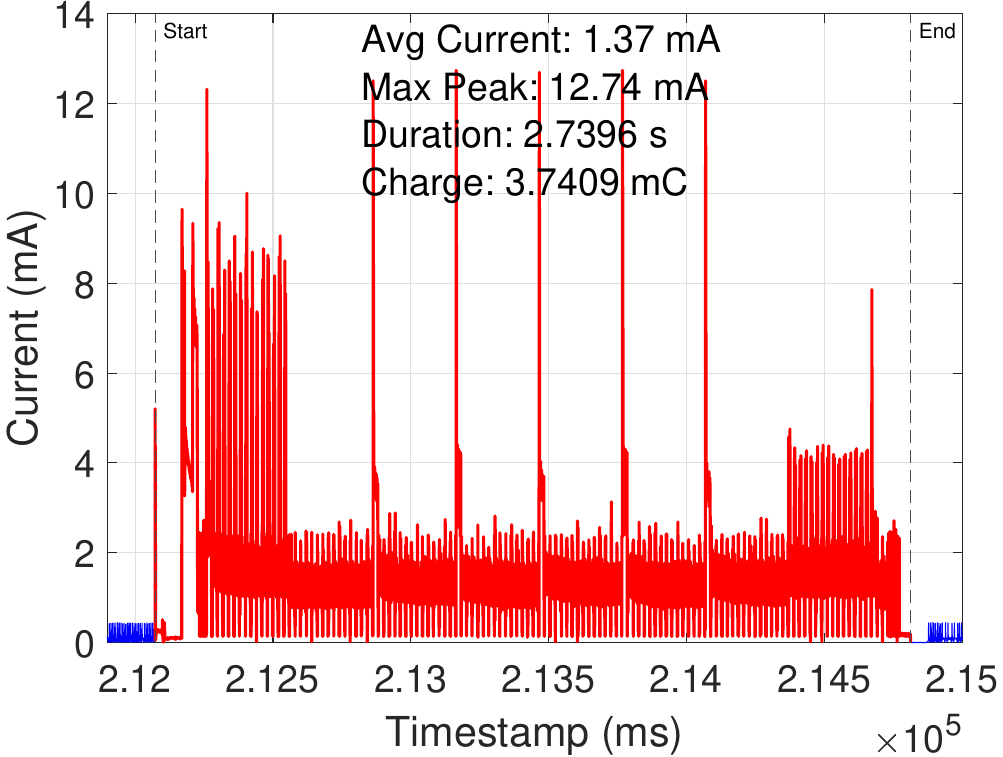}\label{fig:eink}}
\hfil
\subfloat[]{\includegraphics[width=0.51\linewidth]{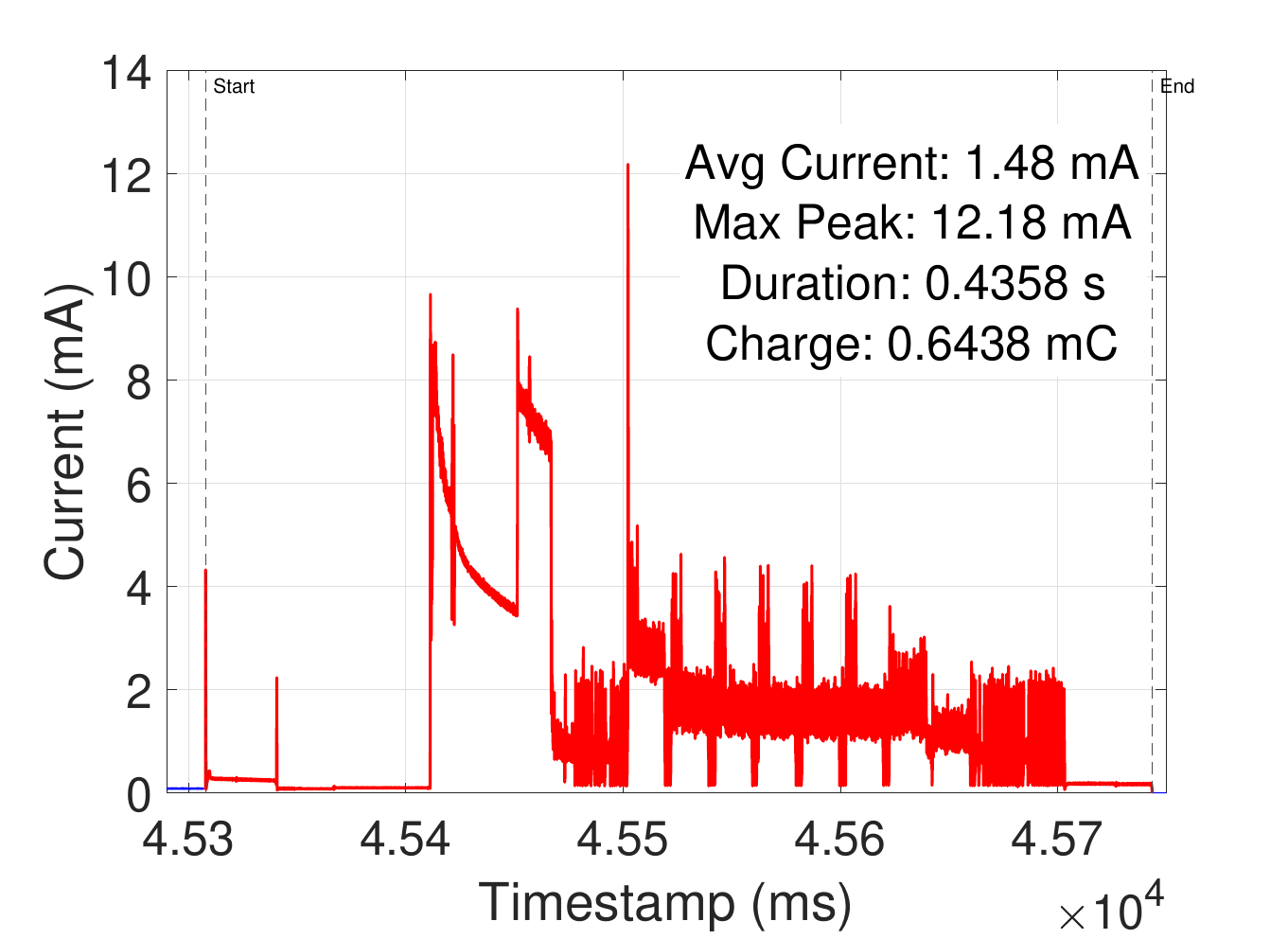}\label{fig:einkopt}}
\caption{Current profiles for (a) Original E-ink display configuration, and (b) Optimized E-ink display configuration.}
\label{fig:einks}
\end{figure}

First, in order to minimize the node's energy consumption, particular attention was given to optimizing the E-ink display operation. The display's firmware Look-Up Table (LUT), or waveform, was refined to achieve lower power consumption during active refresh cycles. In the original configuration, the E-ink display exhibited an average current of approximately 1.4~mA over a 2.8~s refresh period, as illustrated in Fig.~\ref{fig:eink} (when measured exclusively across the E-ink driver component on the node). In the optimized configuration (Fig.~\ref{fig:einkopt}), the average current remains comparable at 1.5~mA; however, the active duration is significantly reduced to approximately 435~ms. Consequently, the display refresh time is reduced by 84\%, leading to a decrease in energy consumption from 12.39~mJ to 2.13~mJ—an overall reduction of approximately 83\%.

\subsubsection{BLE uplink and downlink and VLC uplink only (low-power operation) \label{sec:ble_dlul_vlc_ul}} 
\begin{figure}[t]
\centering
\subfloat[]{\includegraphics[width=0.75\linewidth]{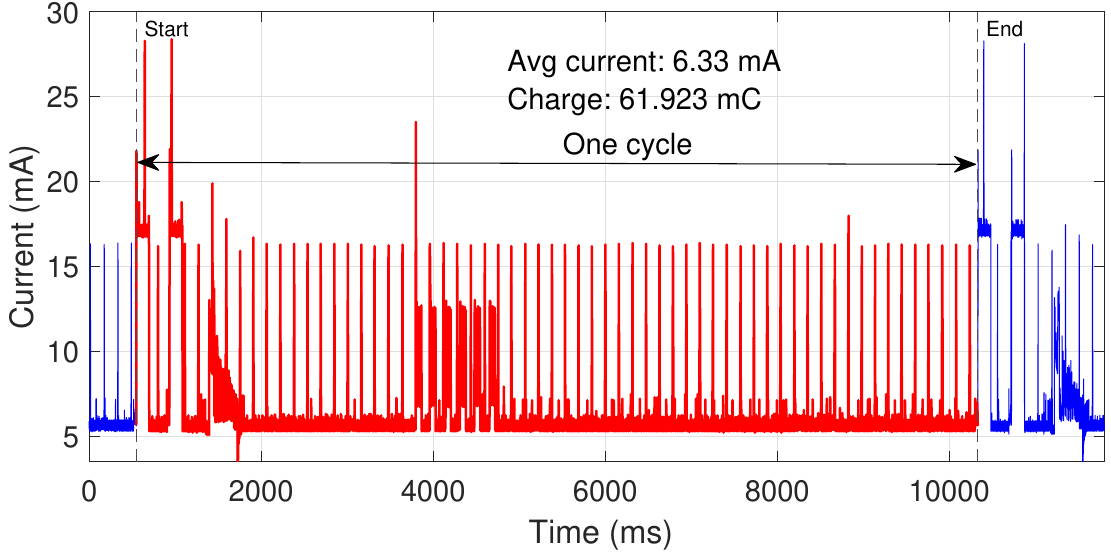}\label{fig:normal}}
\hfil
\subfloat[]{\includegraphics[width=0.75\linewidth]{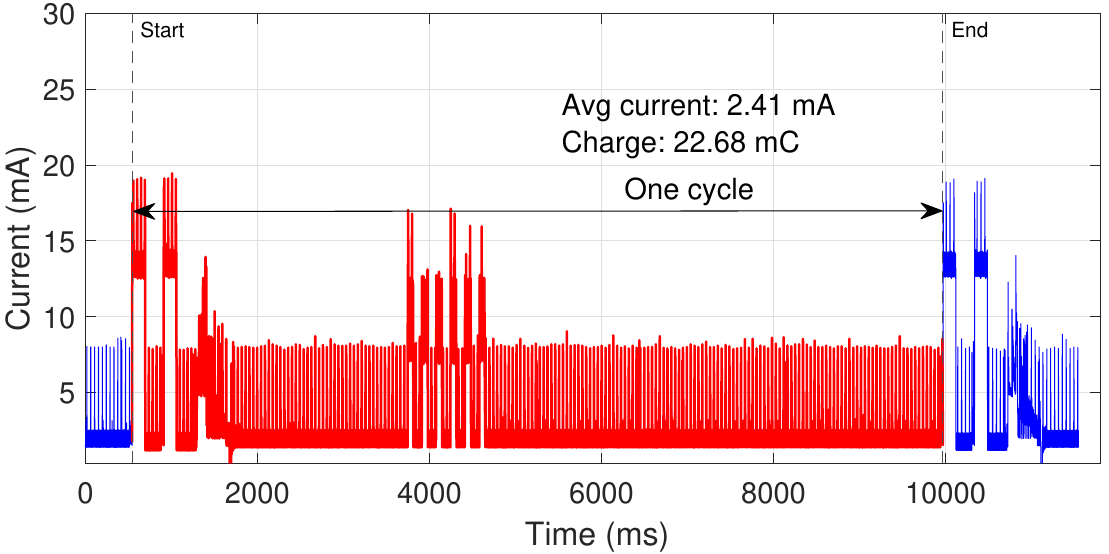}\label{fig:lowpower_opteink}}
\caption{Current profile of node when all functionalities are enabled: (a) Normal, (b) Low-power.}
\label{fig:normalvslowpower}
\end{figure}

Second, software-level optimizations were implemented to reduce the overall current consumption of the RIoT node while performing its core functions. The VLC subsystem, which relies on timers and interrupts to control the PWM signal for optical transmission, increases the computational workload of the BLE SoC (nRF52833). This elevated processing demand results in higher energy consumption due to increased coordination among subsystems.
To minimize power usage, VLC functions—particularly for downlink reception—are disabled in the firmware, while BLE communication remains fully operational. The VLC uplink can still be employed in a duty-cycled manner, where the VLC module is activated immediately prior to transmitting each frame chunk and disabled during inter-chunk intervals, as illustrated in Fig.~\ref{fig:lowpower_opteink}. This approach prevents excessive current draw during idle periods.

For a controlled comparison, both normal and low-power modes follow identical operational cycles, including BLE connection, sensing, E-ink display updates, VLC transmission, and idle periods (Fig.~\ref{fig:normalvslowpower}). Under these conditions, the low-power configuration reduces the average energy consumption per cycle by approximately 63\%, from 0.2043 J in the normal mode (Fig.~\ref{fig:normal}) to 0.0748 J in the low-power mode (Fig.~\ref{fig:lowpower_opteink}).

\begin{figure}[t]
\centering
\subfloat[]{\includegraphics[width=1\linewidth]{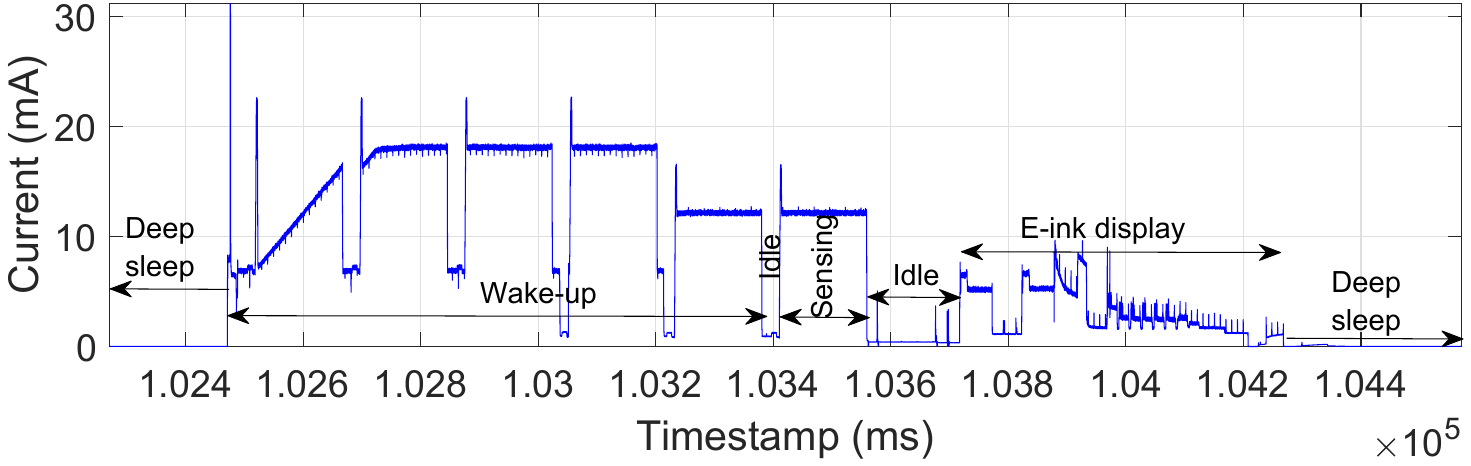}\label{fig:vlp_ble_off_nbvlc_off}}
\hfil
\subfloat[]{\includegraphics[width=1\linewidth]{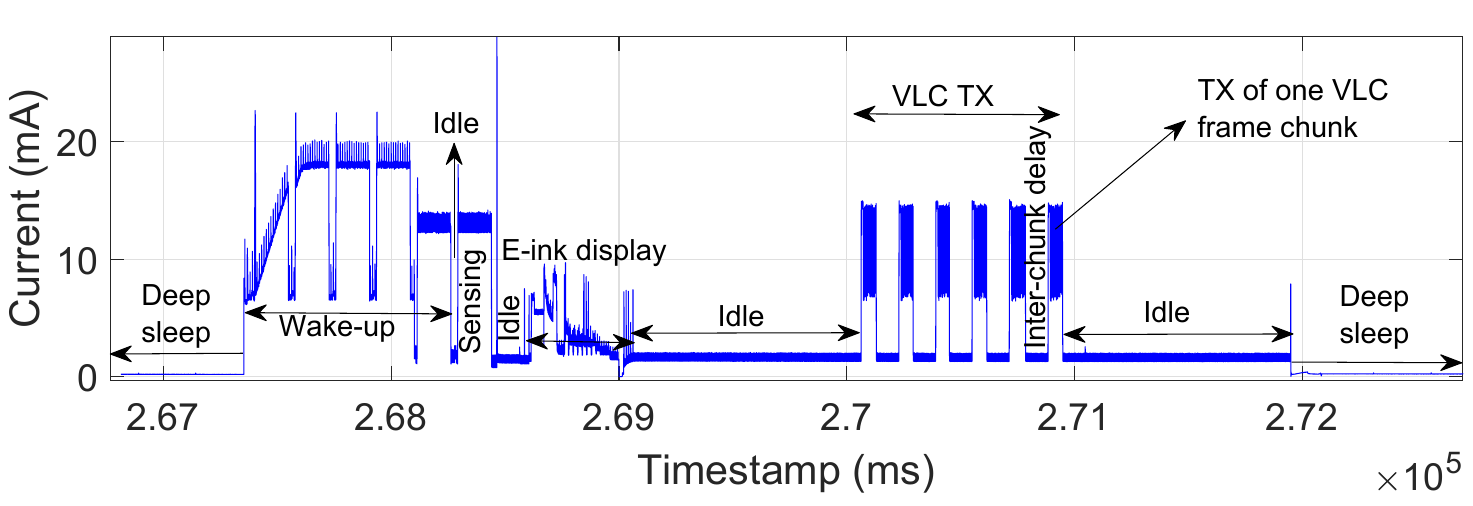}\label{fig:vlp_ble_off_nbvlc_on}}
\caption{Current profile of node during very low-power operation with: (a) both BLE and VLC off, and (b) BLE off and VLC uplink on.}
\label{fig:verylowpower}
\end{figure}

\begin{table}[t]
\centering
\caption{Average current measured during very low-power mode with both VLC and BLE disabled.}
\label{tab:verylowpower}
\setlength{\tabcolsep}{6pt}
\begin{tabular}{@{}p{0.42\columnwidth} p{0.20\columnwidth} p{0.20\columnwidth}@{}}
\toprule
\makecell[l]{\textbf{Node operations}} &
\makecell[c]{\textbf{with VLC}\\\textbf{RX module (mA)}} &
\makecell[c]{\textbf{without VLC RX}\\\textbf{module (mA)}} \\
\midrule
\makecell[l]{Wake-up ($\sim$909\,ms)} & 14.2 & 13.9 \\ \specialrule{0.2pt}{1pt}{1pt}
\makecell[l]{Idle between wake-up and\\ sensing ($\sim$30\,ms)} & 1.34 & 1.02 \\ \specialrule{0.2pt}{1pt}{1pt}
\makecell[l]{Sensing ($\sim$149\,ms)} & 12.5 & 12.2 \\ \specialrule{0.2pt}{1pt}{1pt}
\makecell[l]{Idle between sensing and\\ E-ink display ($\sim$116\,ms)} & 0.843 & 0.435 \\ \specialrule{0.2pt}{1pt}{1pt}
\makecell[l]{E-ink display update ($\sim$544\,ms)} & 3.33 & 2.97 \\ \specialrule{0.2pt}{1pt}{1pt}
\makecell[l]{Deep sleep ($\sim$1:16.1\,min)} & 0.344 & 0.0047 \\
\bottomrule
\end{tabular}
\end{table}

\begin{table}[t]
\centering
\caption{Average current measured during very low-power mode with VLC uplink TX functionality enabled (BLE OFF).}
\label{tab:verylowpower_nbvlc}
\setlength{\tabcolsep}{6pt}
\begin{tabular}{@{}p{0.42\columnwidth} p{0.20\columnwidth} p{0.20\columnwidth}@{}}
\toprule
\makecell[l]{\textbf{Active phase node operations}} &
\makecell[c]{\textbf{with VLC } \\ \textbf{RX module (mA)}} &
\makecell[c]{\textbf{without VLC} \\ \textbf{RX module (mA)}} \\
\midrule
\makecell[l]{Wake-up ($\sim$911\,ms)} & 14.2 & 14.0 \\ \midrule
\makecell[l]{Idle between wake-up and \\sensing ($\sim$30\,ms)} & 1.48 & 1.34 \\ \midrule
\makecell[l]{Sensing ($\sim$149\,ms)} & 12.72 & 12.54 \\ \midrule
\makecell[l]{Idle between sensing and \\E-ink display ($\sim$165\,ms)} & 1.53 & 1.37 \\ \midrule
\makecell[l]{E-ink display update ($\sim$0.45\,s)} & 3.47 & 3.32 \\ \midrule
\makecell[l]{Idle between E-ink and VLC TX \\($\sim$1\,s), \\ Inter-chunk delay \\ ($\sim$100\,ms), \\Idle after VLC TX ($\sim$1\,s)} & 1.50 & 1.40 \\ \midrule
\makecell[l]{Each VLC TX chunk ($\sim$68\,ms)} & 10.70 & 10.20 \\ \midrule
\makecell[l]{Deep sleep ($\sim$1:12.7\,min)} & 0.344 & 0.202 \\ 
\bottomrule
\end{tabular}
\end{table}

The results in Tables \ref{tab:ble_adv_txpower}, \ref{tab:ble_conn_txpower}, and \ref{tab:normal_lp_firmware01} further confirm that the software optimizations consistently reduce the average current consumption by approximately 4 mA across various operating states. 

During BLE uplink transmission in the low-power configuration, the current draw ranges from 5.91 mA at 0 dBm to 10.18 mA at +8 dBm, over a duration of approximately 3.13 ms. Assuming a connection interval of 45 ms and a supply voltage of 3.3 V, the energy per BLE uplink transmission is approximately 61 µJ, representing a reduction of $\approx$35\% compared to the normal mode.

For VLC uplink transmission segmented into six frame chunks, the low-power mode exhibits an average current of 4.9 mA over 0.91 s, corresponding to a total energy expenditure of $\approx$15 mJ, which is approximately 30\% lower than in the normal mode. These results demonstrate that software-level optimizations can substantially reduce energy consumption while maintaining functional performance across both BLE and VLC subsystems.

\subsubsection{BLE and VLC both OFF (very low-power operation)}
\label{sec:ble_off_vlc_off}
In this configuration, both communication interfaces are disabled, meaning the node neither transmits data to nor receives commands from the AP. The node operates solely in a standalone mode, performing local sensing and actuation tasks such as E-ink display updates, followed by extended periods of deep sleep to minimize energy consumption, as illustrated in Fig.~\ref{fig:vlp_ble_off_nbvlc_off}. 
To further reduce quiescent power consumption during deep sleep, the hardware link connecting the E-ink display’s power-control circuitry to the always-on supply was opened. This modification electrically isolates the display's power-control network during deep sleep, eliminating leakage currents from that stage. Importantly, it does not interrupt the main logic power or display-driving capability, allowing full refresh functionality to be retained while achieving lower standby current. 
The corresponding average current consumption for each operation in this configuration is detailed in Table~\ref{tab:verylowpower}.
We further evaluate the impact of isolating the VLC receiver circuit from the main \textit{VDD} rail on the node’s average current consumption. 
In practice we will not physically cut the VLC RX module. The deactivation of the VLC RX component can be achieved by soldering in a 0 Ohm resistor to allow powering of the circuitry by switching the relevant nRF52833 General-Purpose Input/Output (GPIO) pin to HIGH during normal operation, and to shut down the circuit by switching the pin to LOW during sleep, thereby enhancing energy efficiency. 
As shown in Table~\ref{tab:verylowpower}, when the VLC receiver module is excluded from the circuit, the deep-sleep current decreases markedly—from approximately 344~\textmu A to only 5~\textmu A—corresponding to an estimated 99\% reduction. Slight decrease in current consumption is also noted across the other states. These results highlight the significant energy-saving benefits achievable through simple hardware-level modifications.

\subsubsection{BLE OFF and VLC uplink ON}
This operational mode is similar to that described in Section~\ref{sec:ble_off_vlc_off}, with the addition of VLC uplink transmission to the AP during the active phase, as illustrated in Fig.~\ref{fig:vlp_ble_off_nbvlc_on}. The BLE module remained disabled throughout this test. Compared to the profile in Fig.~\ref{fig:vlp_ble_off_nbvlc_off}, an additional idle period of approximately 1~s was inserted after the E-ink display update and VLC transmission phase. The measured average current for each operation is summarized in Table~\ref{tab:verylowpower_nbvlc}.
Consistent with the results in Table~\ref{tab:verylowpower}, the exclusion of the VLC receiver module again yields a reduction in deep-sleep current—from approximately 344~\textmu A to 202~\textmu A—representing a decrease of about 41\%. Minor reductions are also observed in other operational states. However, the overall improvement is less pronounced than in Table~\ref{tab:verylowpower}, primarily due to the additional computational overhead introduced by the VLC transmission functions in the firmware (as previously investigated in Section \ref{sec:ble_dlul_vlc_ul}), which increase CPU activity and coordination among subsystems, thereby elevating energy consumption.

\subsection{Digital Twin Evaluation \label{sec:Evaluation_DT}} 

To assess the effectiveness of the developed DT and EUNO algorithm, we implemented EUNO within the DT framework and compared its performance against a baseline method, the Energy-aware Threshold-based Node Optimization (ETNO) algorithm.  The ETNO algorithm optimizes the configuration of each RIoT node based solely on the current energy level, adjusting its operational state and communication modality once predefined thresholds are reached. Specifically, ETNO defines two key thresholds: a \textit{sleep threshold}, representing the energy level below which the node automatically transitions into sleep mode, and a \textit{conservation threshold}, below which a node operating in performance mode switches to conservation mode, and above which a node resumes normal functionality. 

\begin{figure}[t!]
	\centering
		\scalebox{0.92}{\includegraphics[width=0.5 \textwidth]{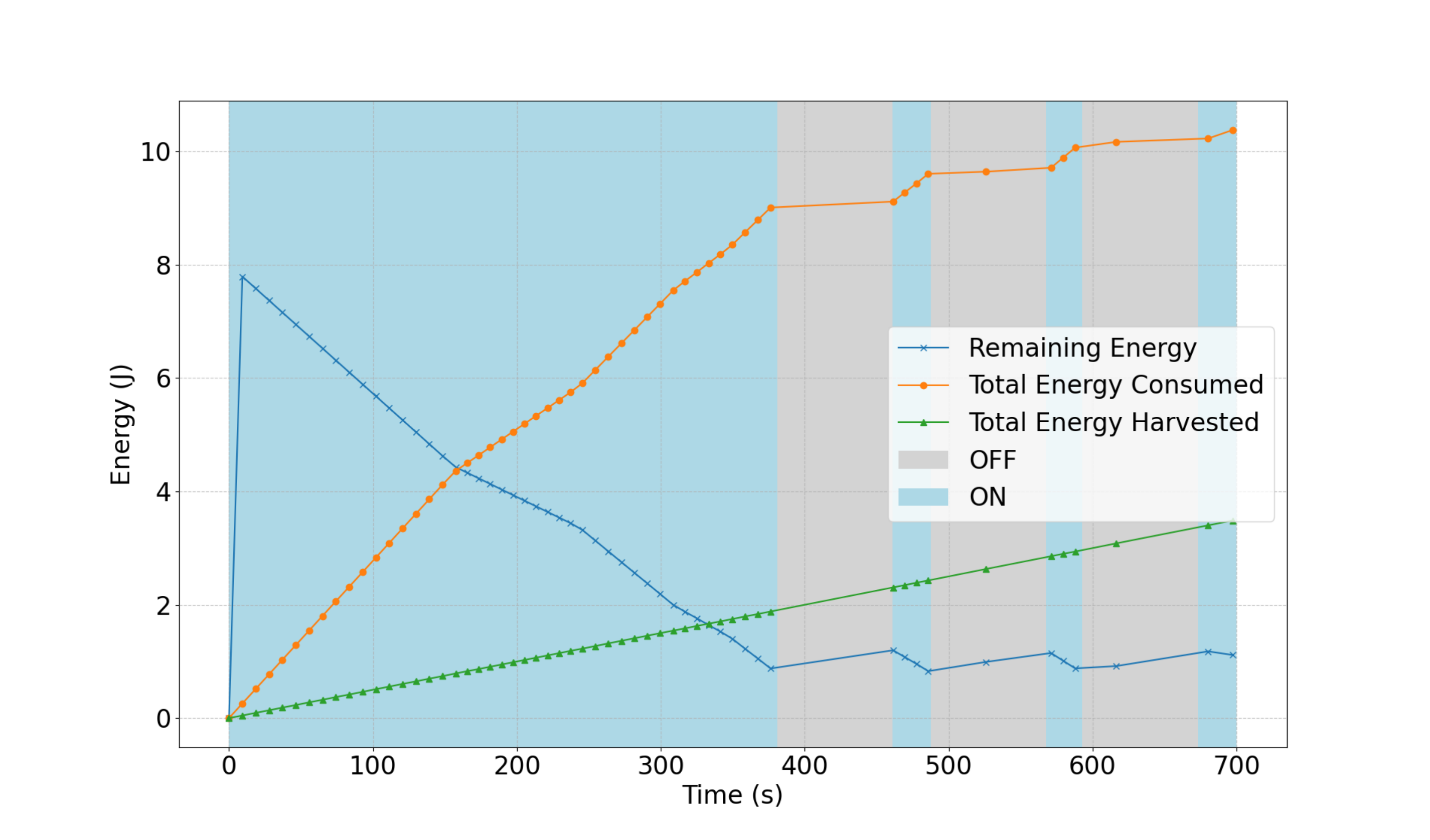} } 
	\caption{ Energy consumption, harvested energy, and remaining energy of a RIoT node using only OWC over time. Blue regions indicate performance mode, and gray regions indicate sleep mode. 
    } 
	\label{fig:OWC_Sim}
\end{figure}

First, Figure~\ref{fig:OWC_Sim} illustrates an example of the evolution of energy consumption, harvested energy, and remaining energy over time for a RIoT node. In this example, only a sleep threshold is enforced, considering OWC as the sole transmission modality. The aim of this figure is to demonstrate the importance of adaptive mode transitions. Specifically, switching the RIoT node from standard transmission to a sleep state allows for preventing complete energy depletion while enabling effective energy harvesting during low-activity periods. 

\begin{figure}[t]
\centering
\subfloat[]{\includegraphics[width=0.9\linewidth]{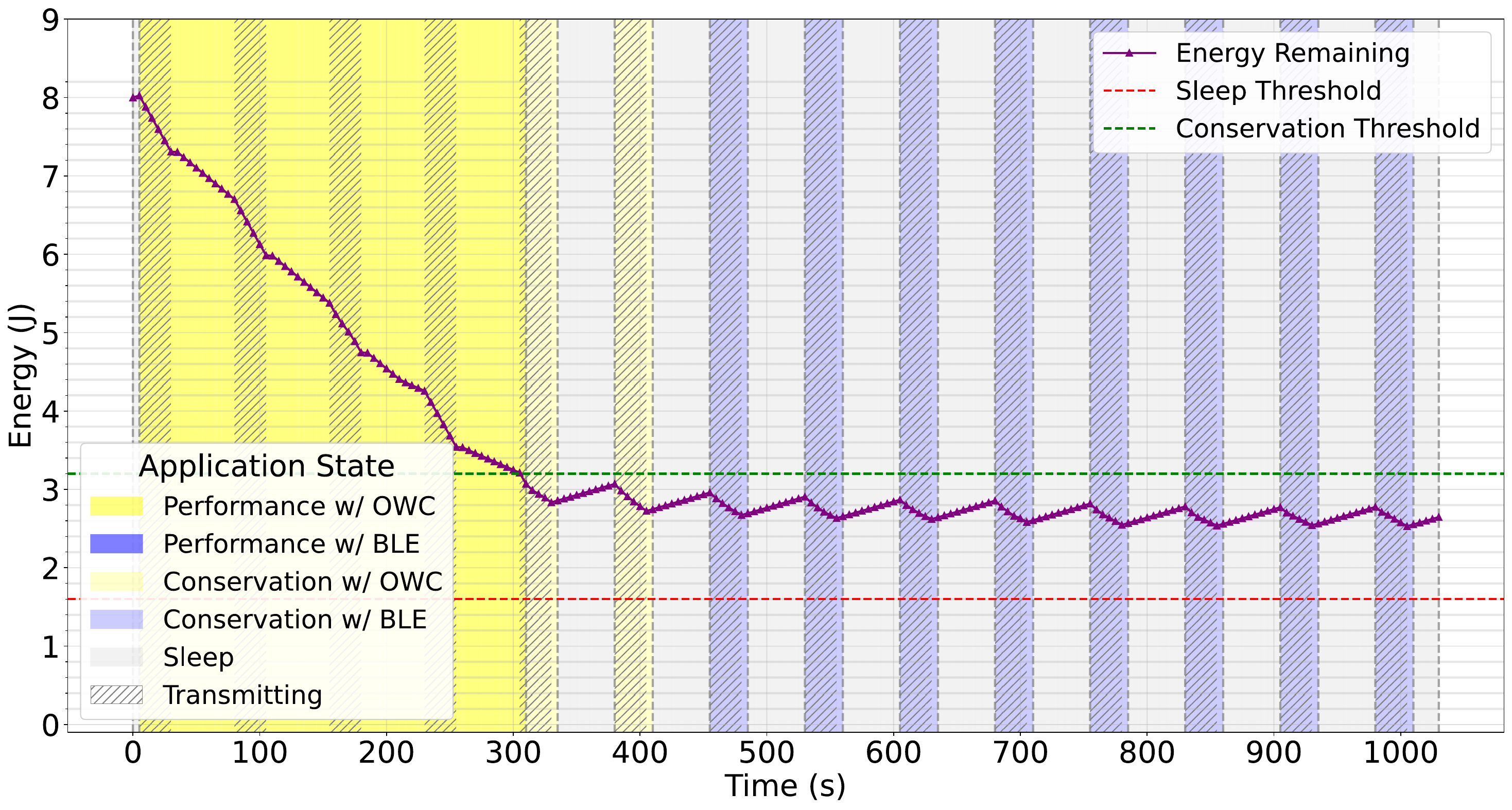}\label{fig:ETNO-no-itsp}}
\hfil
\subfloat[]{\includegraphics[width=0.9\linewidth]{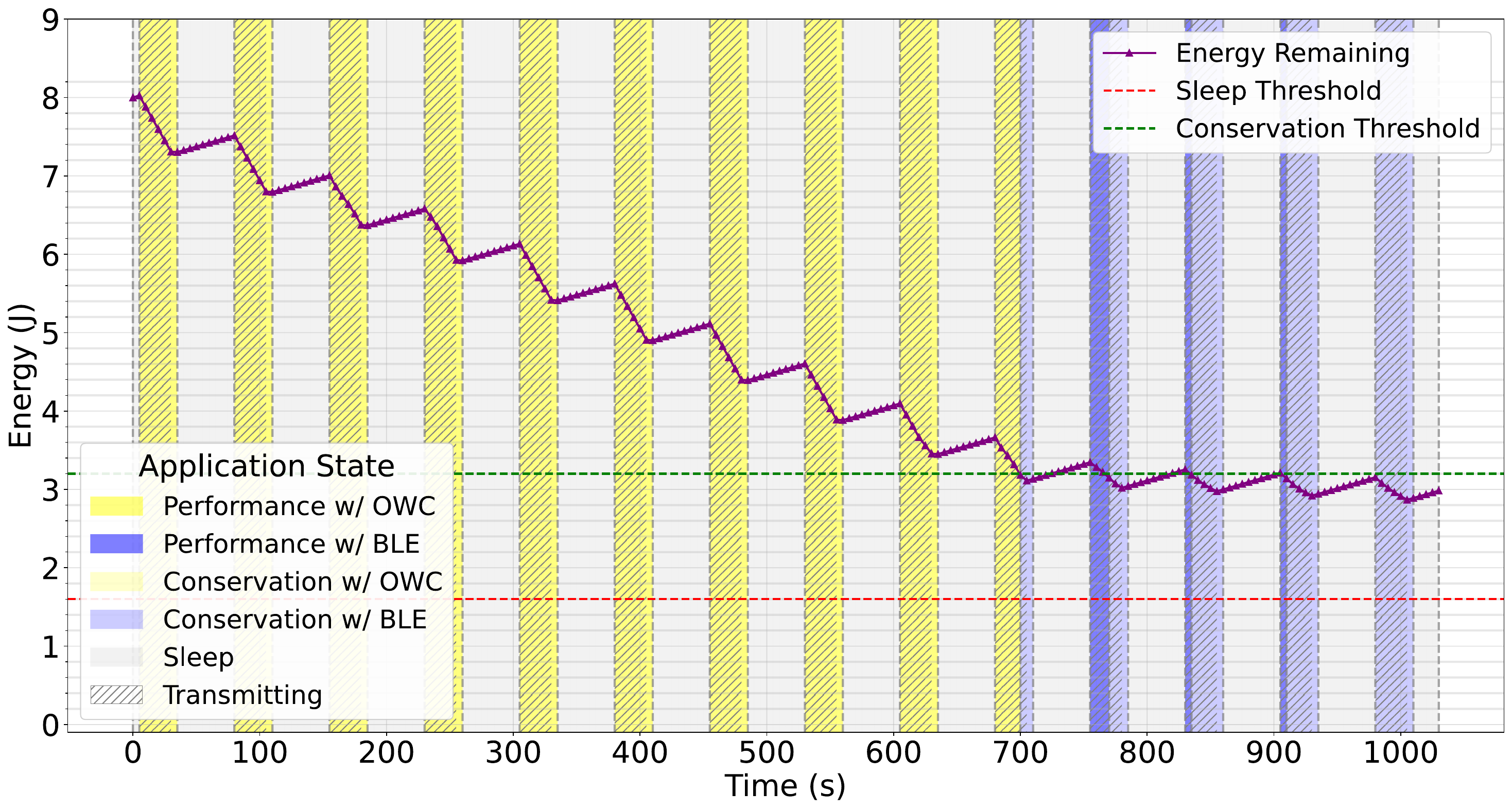}\label{fig:ETNO-with-itsp}}
\caption{Evolution of the remaining energy over time for a RIoT node at 300 kb/s target data rate, under the ETNO algorithm: (a) without inter-transmission sleep periods and (b) with inter-transmission sleep periods.}
\label{fig:ETNO-itsp-compare}
\end{figure}

\begin{figure}[t]
\centering
\subfloat[]{\includegraphics[width=0.9\linewidth]{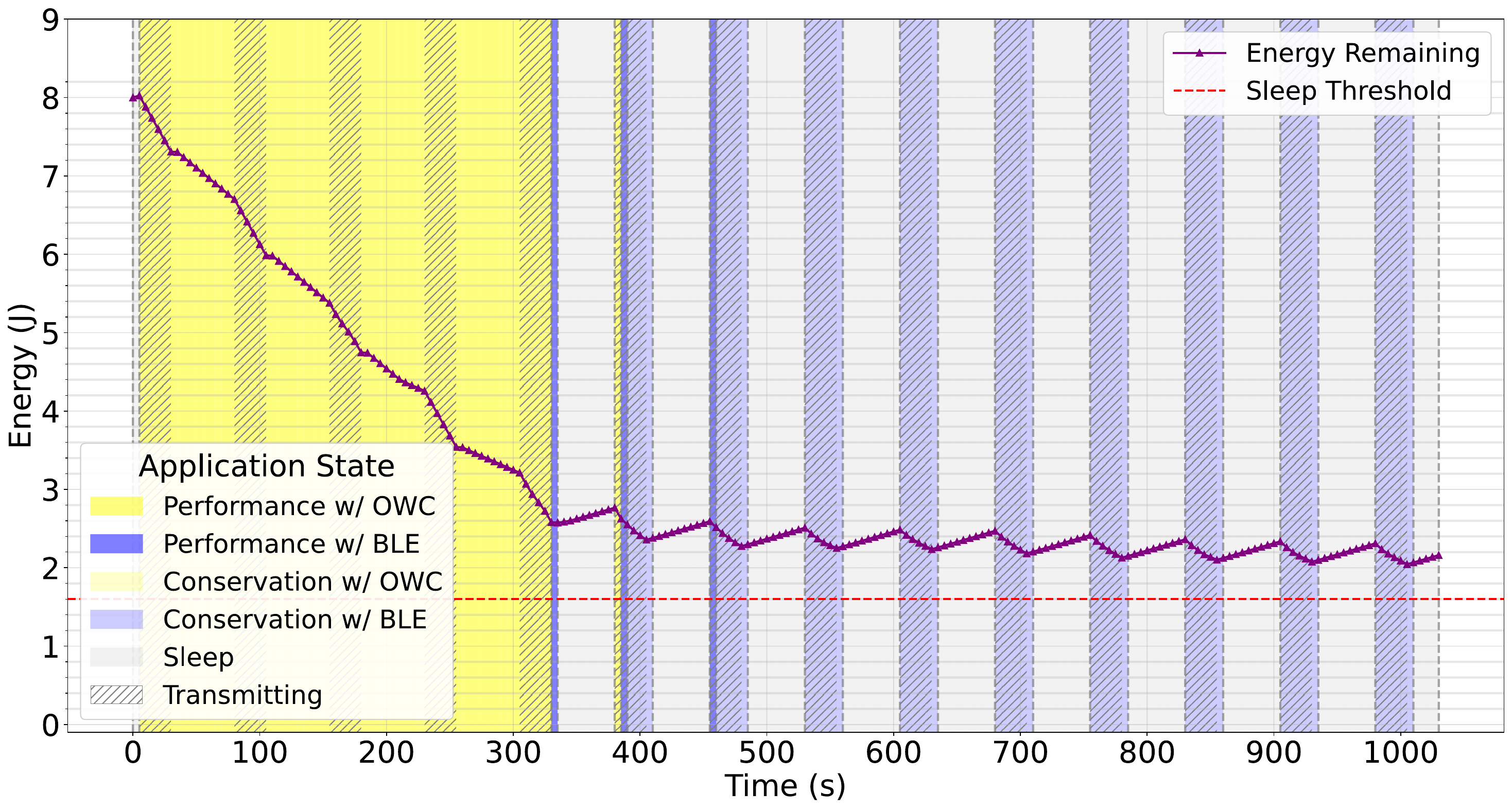}\label{fig:EUNO-no-itsp}}
\hfil
\subfloat[]{\includegraphics[width=0.9\linewidth]{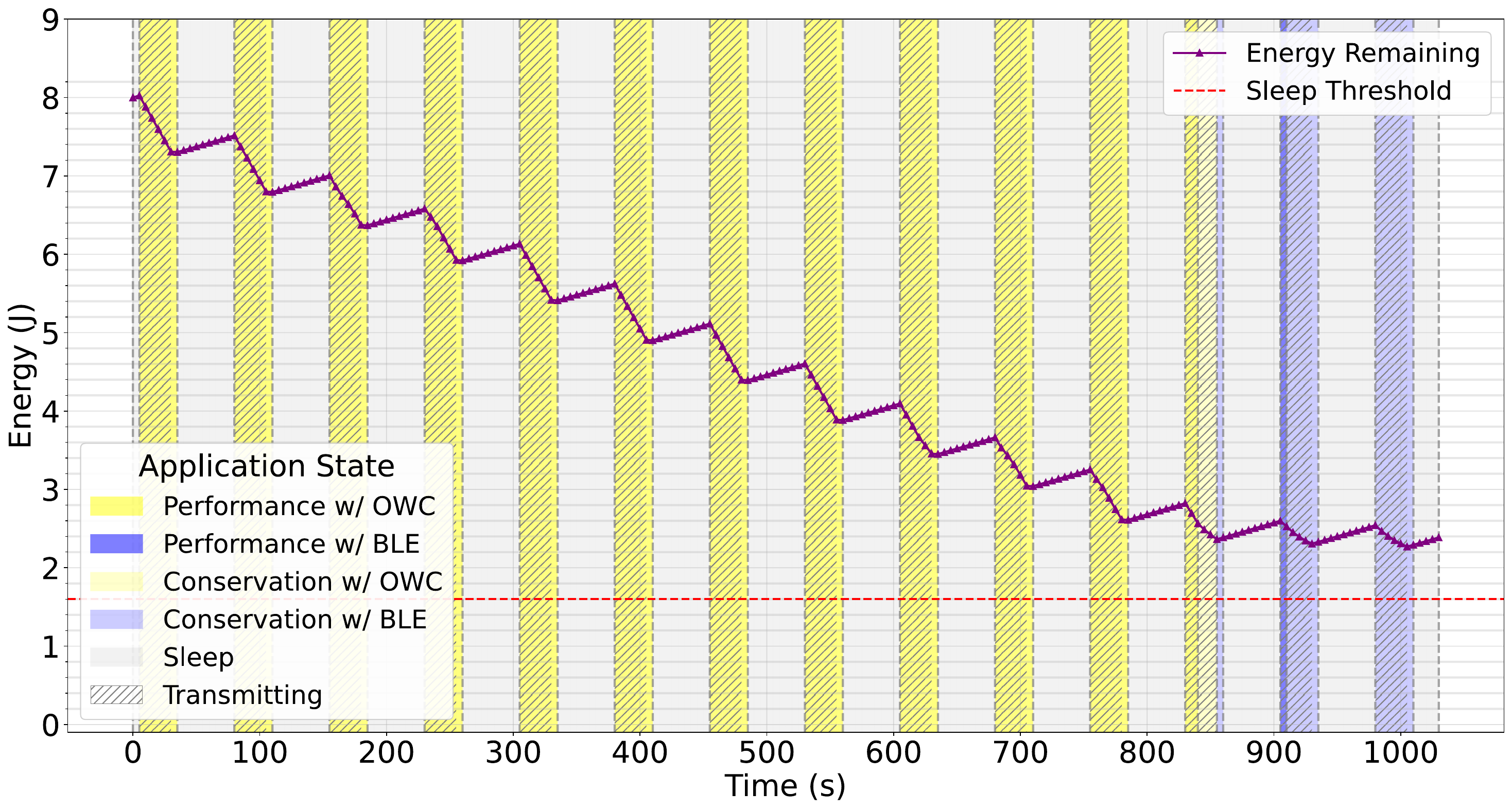}\label{fig:EUNO-with-itsp}}
\caption{Evolution of the remaining energy over time for a RIoT node at 300 kb/s target data rate, under the EUNO algorithm: (a) without inter-transmission sleep periods and (b) with inter-transmission sleep periods.}
\label{fig:EUNO-itsp-compare}
\end{figure}

Second, Figures~\ref{fig:ETNO-itsp-compare} and \ref{fig:EUNO-itsp-compare} present a comparative evaluation of the ETNO and EUNO algorithms under two operational scenarios: (a) without inter-transmission sleep periods and (b) with inter-transmission sleep periods. Figure~\ref{fig:ETNO-itsp-compare} depicts the evolution of the remaining energy for a RIoT node managed by the ETNO algorithm, while Figure~\ref{fig:EUNO-itsp-compare} shows the corresponding results for a node operating under the proposed EUNO algorithm. This comparison highlights the impact of each optimization approach on the node's energy dynamics, revealing how adaptive EUNO enhances energy sustainability and operational efficiency compared to the ETNO.  
Each simulation run lasted 1,025 seconds, with an additional 5-second initialization period for application startup. The network topology consisted of three end nodes, each equipped with a 8-joule battery. Packet transmissions were scheduled using a round-robin approach, with a 25-second transmission period per node. Each application generated 512-byte packets at an application data rate of 300 kb/s, which was reduced to 60 kb/s when operating in conservation mode. Node-specific metrics, including the remaining energy and total transmitted data, were monitored for the first end node in the topology. In the ETNO algorithm, the conservation threshold was set to 40\% of the total energy capacity, while the sleep threshold was defined at 20\%. 

In Figure~\ref{fig:ETNO-itsp-compare}-(a), the node remains continuously active between transmissions, leading to a rapid and nearly linear decline in its remaining energy. This continuous activity causes the node to deplete its energy reserves quickly, forcing an early transition to the conservation state and thereby reducing its overall operational performance. Conversely, Figure~\ref{fig:ETNO-itsp-compare}-(b) demonstrates that introducing short sleep periods between transmissions substantially mitigates energy depletion. During these intervals, the node enters a sleep mode, allowing it to harvest energy and partially recover its reserves, resulting in a more balanced energy trajectory and extended lifetime. 
As observed in both cases, the node initially employs OWC for data transmission, as it provides the best SNR under the given conditions. When the remaining energy falls below the conservation threshold, the node automatically switches both its operational mode (from Performance to Conservation) and its communication modality (from OWC to BLE) to preserve power (see Figure~\ref{fig:ETNO-itsp-compare}). Nonetheless, in the first case, i.e., without inter-transmission sleep phases, the node's higher initial energy consumption causes it to reach the sleep threshold much earlier, leading to reduced overall performance. 
The continuously active node transferred a total of 5.78 MB of data, with an average rate of 132.1 kb/s, while the node incorporating inter-transmission sleep periods achieved 10.72 MB, with an average rate of 245.1 kb/s (it should be noted that the effective data rates were calculated in relation to the time the node was allowed to transmit and not the whole simulation time). This confirms that periodic sleep cycles, coupled with dynamic mode and modality switching, substantially enhance both energy efficiency and data throughput in RIoT operations. 

Compared to the ETNO results, the EUNO algorithm exhibits a smoother and more balanced energy consumption trajectory, maintaining higher energy levels for longer periods (see Figure~\ref{fig:EUNO-itsp-compare}). This improvement stems from EUNO's utility-based decision mechanism, which adaptively determines the optimal operational mode and transmission modality based on the node's residual energy, predicted consumption, and communication performance. Interestingly, under EUNO, the nodes tend to switch from performance mode to conservation mode even changing from OWC to BLE communication. This proactive adaptation enables the nodes to achieve higher transmission rates before reaching the low-power threshold. Consequently, the node operating without inter-transmission sleep periods transmitted a total of 6.68 MB of data, with an average data rate of 152.6 kb/s, while the node with inter-transmission sleep periods achieved 11.32 MB, with an average data rate of 258.9 kb/s. 

\begin{figure}[t]
\centering
\subfloat[]{\includegraphics[width=0.9\linewidth]{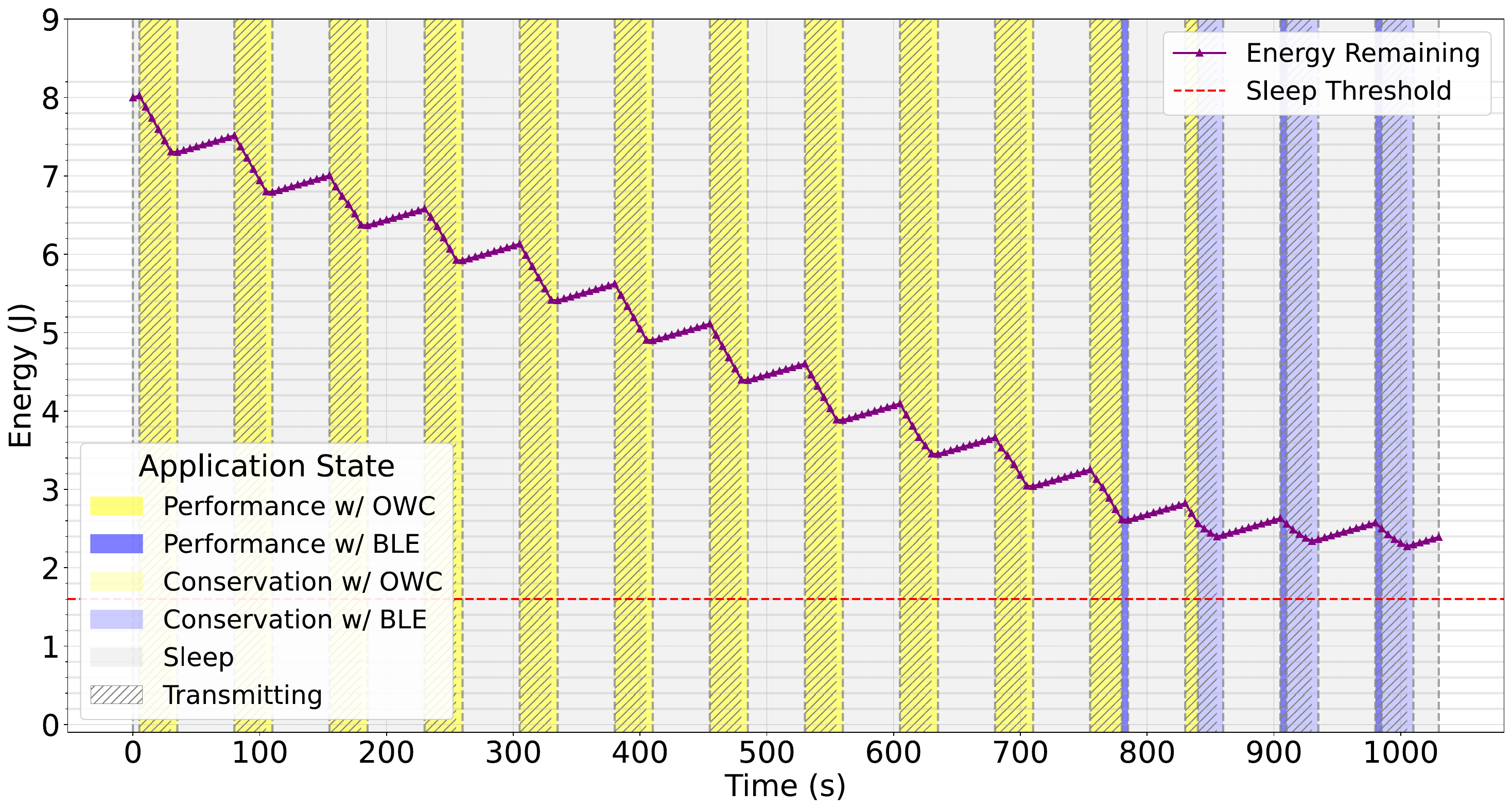}\label{fig:no-osc-prot}}
\hfil
\subfloat[]{\includegraphics[width=0.9\linewidth]{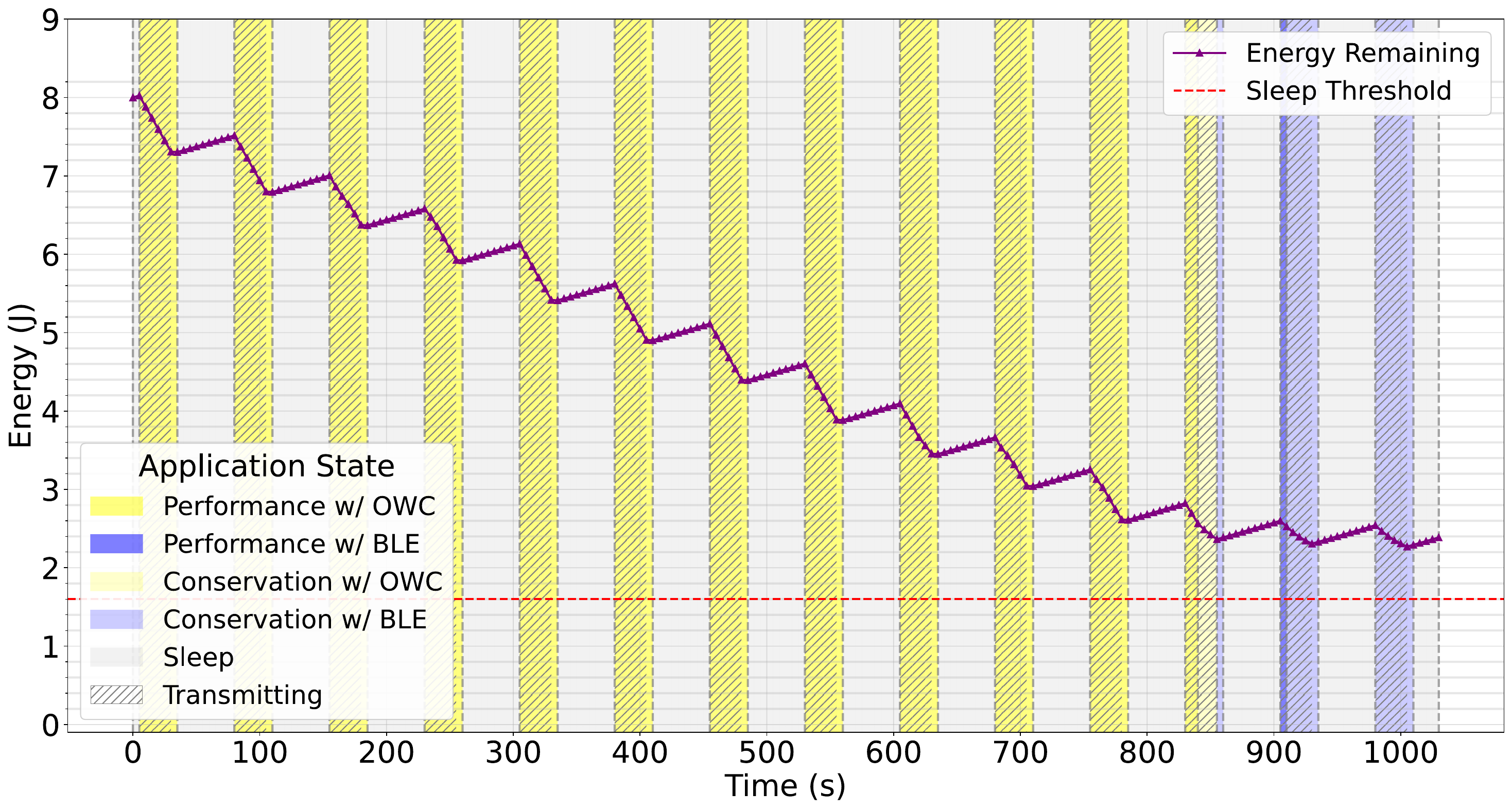}\label{fig:osc-prot}}
\caption{Evolution of the remaining energy over time for a RIoT node at 300 kb/s target data rate, under the EUNO algorithm: (a) without modality oscillation protection, and (b) with modality oscillation protection.}
\label{fig:osc-prot-compare}
\end{figure}

To mitigate the frequent oscillations between communication modalities, as observed in Fig.~\ref{fig:osc-prot-compare}(a), a switching penalty $p_{ch}$ was incorporated into the EUNO algorithm. Interestingly, by tuning this penalty using historical data, the algorithm effectively discourages unnecessary switching without significantly impacting performance. As shown in Fig.~\ref{fig:osc-prot-compare}(b), this leads to more stable and deliberate modality selection, with a clear reduction in switching events.

\begin{figure}[t!]
	\centering
		\scalebox{1.4}{\includegraphics[width=0.35 \textwidth]{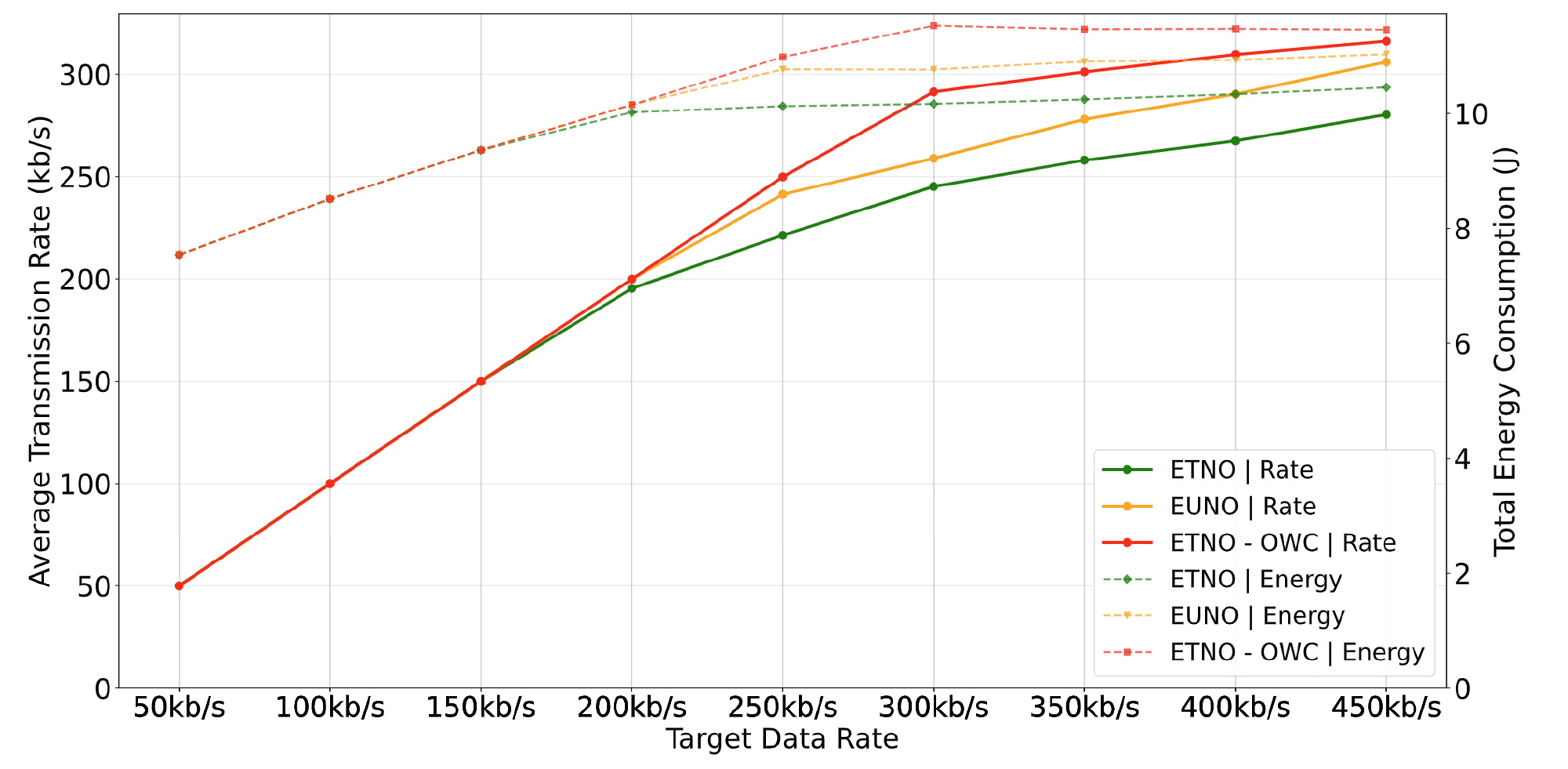} } 
	\caption{ Average node transmission data rate as a function of the target application data rate for ETNO, ETNO-OWC, and EUNO algorithms.   } 
	\label{fig:ETNO_EUNO_datarate}
\end{figure}

Finally, the aforementioned observations are further supported by the results in Fig.~\ref{fig:ETNO_EUNO_datarate}, which presents the average transmission data rate achieved under different target application data rates. In this comparison, the EUNO algorithm is evaluated against ETNO and ETNO-OWC. The latter applies the ETNO strategy but constrained to OWC only. At low target data rates, all algorithms perform similarly, as intermittent sleep periods allow nodes to recover sufficient energy. However, as the target rate increases, energy stress begins to impact node operation. In these conditions, EUNO clearly outperforms ETNO and ETNO-OWC by maintaining a higher effective data rate while preserving energy consumption. This improvement stems from EUNO’s ability to dynamically adapt both operational mode and communication modality based on residual energy and environmental conditions, achieving a more efficient trade-off between performance and energy sustainability.  

\section{Conclusion \label{sec:conclusion}}

This paper presented a RIoT framework that enables dynamic integration of optical and radio wireless communication, allowing both nodes and access points to operate in RF, OWC, or hybrid modes according to channel conditions, application requirements, and energy availability. By combining hybrid communication, real hardware-calibrated energy models, and a high-fidelity DT, the proposed system offers a scalable and realistic platform for evaluating sustainable 6G IoT networks. 
A proactive, low-complexity cross-layer optimization algorithm (EUNO) was also introduced. Using a unified utility function, EUNO enables energy-aware and context-driven adaptation of communication modality and operating mode, while remaining suitable for resource-constrained hardware. 
Experimental and DT-based results show that EUNO outperforms threshold-based baselines, achieving approximately 6\% higher average data rate while maintaining energy sustainability and reducing unnecessary switching between modalities. These results demonstrate the effectiveness of combining DT technology with hybrid RF/OWC communication for resilient and energy-efficient IoT systems. 

\balance 

\bibliographystyle{IEEEtrannames}
\bibliography{Ref}

\end{document}